\documentclass{iopjournal-clean}

\usepackage{amsmath}
\usepackage{amsfonts}
\usepackage{hyperref}
\usepackage{graphicx}
\usepackage{xcolor}
\usepackage{relsize}
\usepackage{ragged2e}
\usepackage{upgreek}

\DeclareMathAlphabet{\mathpzc}{OT1}{pzc}{m}{it}

\newcommand{\properbinom}[2]{\left(\!{{#1}\atop{#2}}\!\right)}
\newcommand{\expe}{\mathrm{e}}
\newcommand{\imagi}{\mathrm{i}}

\usepackage[backend=biber,citestyle=authoryear,bibstyle=authortitle,natbib,hyperref=true]{biblatex}
\newcommand{\dlmf}[1]{\citep[\S #1]{DLMF}}
\newcommand{\bmp}[1]{\citep[\S #1]{BMP}}
\newcommand{\gradsh}[1]{\citep[Eq.~#1]{GR}}
\newcommand{\prud}[1]{\citep[#1]{PB}}
\newcommand{\mf}[1]{\citep[#1]{MF}}
\newcommand{\coordref}[2]{~~{\footnotesize\normalfont (Morse \& Feshbach #1; Miller #2)}}

\newcommand{\refedit}[1]{#1}

\renewcommand\vec{\mathbf}
\newcommand{\intddd}{\ensuremath{\int_{\mathbb{R}^3} \! d^{3\,}\!\vec{r}}}

\newcommand{\unitvector}{\hat{\vec r}}

\newcommand{\grad}{\nabla}

\newcommand{\dop}{\ensuremath{\mathcal{D}}}

\newcommand{\pop}{\ensuremath{\mathcal{P}}}
\newcommand{\kop}{\ensuremath{\mathcal{K}}}
\newcommand{\jop}{\ensuremath{\mathcal{J}}}

\newcommand{\Lap}{\ensuremath{\nabla^2}}

\newcommand{\eivec}{\ensuremath{\lambda}}

\DeclareMathOperator{\cosech}{csch}
\DeclareMathOperator{\cosec}{csc}
\DeclareMathOperator{\arsinh}{arsinh}
\DeclareMathOperator{\sech}{sech}

\DeclareMathOperator{\ce}{ce}
\DeclareMathOperator{\se}{se}
\DeclareMathOperator{\me}{me}

\newcommand{\hypergeom}[5]{\ensuremath{ {}_{#1} F_{#2}\!\left(\left.\!%
     \begin{matrix}
       #3 \\
       #4
     \end{matrix}%
   \: \!\right|\! #5 \!\right)}}

\begin{document}

\articletype{Paper} %

\title{Basis sets and Coulomb resolutions in rotational coordinates}

\author{Edward Lilley\orcid{0000-0003-0011-4668}}

\email{edward.lilley@univie.ac.at}

\affil{Department of Astrophysics, University of Vienna, T\"urkenschanzstra{\ss}e 17, 1180 Vienna, Austria}

\keywords{coulomb resolution, basis functions, separation of variables}

\begin{abstract}
  Using generalised Laplacian symmetry operators, we construct basis
  sets or Coulomb resolutions in several separable coordinate systems,
  including two $R$-separable systems. This expands the possible
  geometries in which basis set construction is feasible, a problem
  which is relevant to both galactic dynamics and computational
  chemistry. In particular we derive three basis sets (two in prolate
  spheroidal and one in cylindrical coordinates) which are expressible
  in closed-form using a single Jacobi polynomial. We also show how
  any spherical polar or prolate spheroidal basis set may be
  transformed into a bispherical or toroidal basis set.
\end{abstract}

\justifying

\section{Introduction}\label{sec:intro}

We define the energy (or `energetic') inner product between two
mass or charge densities as\footnote{Using Green's identities and
  Poisson's equation $\Lap \Phi = 4\pi \varrho$ this can also be
  written as
\[
  \langle\varrho_1, \varrho_2\rangle = \frac{1}{4\pi} \intddd \: \grad \Phi_1 \cdot \overline{\grad \Phi_2}
  = -\intddd \: \varrho_1 \overline{\Phi_2},
\]
\refedit{so it can also be interpreted (on potential functions $\Phi$)
  as the inner product associated with the homogeneous Sobolev
  semi-norm of order $1$.}
} %
\begin{equation}\label{eq:self}
  \langle\varrho_1, \varrho_2\rangle = \intddd \! \intddd^\prime \: \frac{\varrho_1(\vec{r}) \overline{\varrho_2(\vec{r}^\prime)}} {\lVert \vec{r} - \vec{r}^\prime\rVert}. %
\end{equation}
In the galactic dynamics literature
\citep{CB72,CB73,Ka76,We99,Lilley2023} a \textit{basis
  set}\footnote{\refedit{Commonly also called a \textit{biorthogonal}
    or \textit{biorthonormal} basis set. This terminology is
    mathematically imprecise, but we can write
    $\langle\varrho_n, \varrho_m\rangle = -(\varrho_n, \Phi_m)$, which
    resembles the formal definition of biorthogonality \citep[\S
    1.4]{Higgins1977}; however the two sequences $\{\varrho_n\}$ and
    $\{\Phi_n\}$ are not independent.}} is a set of mass densities
$\{\varrho_n\}$ that obeys the orthogonality relation
\begin{equation}\label{eq:orth}
  \langle \varrho_n, \varrho_m \rangle = N_{n} \: \delta_{nm}
\end{equation}
where $n$ and $m$ are suitable multi-indices and $N_{n}$ a
normalisation constant. In the computational chemistry literature the
`resolution of the Coulomb operator' (hereafter \textit{Coulomb resolution})
was introduced by \citet{Varganov2008}\footnote{Continued in a series
  including
  \citet{Gill2009,Limpanuparb2009,Limpanuparb2011,Limpanuparb2012}.}.
This is a weakly-convergent series representation of the Coulomb
potential of the form
\begin{equation}\label{eq:resolution}
\frac{4\pi}{\lVert \vec{r} - \vec{r}^\prime \rVert} = \sum_{n} \Phi_{n}(\vec{r}) \: \overline{\Phi_{n}(\vec{r}^\prime)} / N_{n}.
\end{equation}
Evidently this is formally equivalent to the galactic dynamicists'
basis set under the identification $\Lap \Phi_n = 4 \pi \varrho_n$. It
is therefore of general interest to find examples of orthogonal sets
$\{ \varrho_n \}$ or $\{ \Phi_n \}$ that are easily computable and
suitable for different applications.

\subsection{\refedit{Applications}}

\refedit{In the context of galactic dynamics, it is typically
  desirable} for the orthogonal density basis functions to resemble,
as far as possible, the underlying physical (unperturbed or
equilibrium) density, so that the resulting expansions can be usefully
truncated at low order. There have been numerous attempts over the
years to construct such functions in a flexible way\footnote{To the
  references listed earlier we can add
  \citet{Sa93,Ea96,Ro96,Qi93,Bro98,Ra09}, among others. There have
  also been various other efforts in both the chemistry and
  astrophysics literature towards generating (not necessarily
  orthogonal) basis functions using differential operators, satisfying
  desirable properties other than those considered here,
  e.g.~\citet{Niukkanen1984,Le79,deZ88} (in general the choice of
  operator is dependent on the family of basis functions under
  consideration).}.  \refedit{%
  In `$N$-body' simulations, where a cloud of $N$ point masses evolve
  dynamically in their own gravitational field,
  one is interested in finding a smooth approximation to the
  gravitational field in a way that avoids direct summation over all $N^2$
  pairs of particles. If the $j$-th particle has mass $\mu_j$ and position $\vec{r}_j$,
  then using a suitable basis set we can approximate the gravitational
  potential as
  \begin{equation}
    \Phi(\vec{r}) \approx -\sum_{n=0}^{n_\mathrm{max}} \sum_{j=1}^N \mu_j \: \overline{\Phi_n(\vec{r}_j)} \: \Phi_n(\vec{r}) / N_n,
  \end{equation}
  which has only $\sim \! Nn_\mathrm{max}$ terms. The potential can be
  re-calculated at every time step in order to evolve the particles'
  dynamics self-consistently \citep{HO92}; or it can be used as a
  convenient time-evolving representation (with `frozen' dynamics) of
  a galaxy or dark-matter halo in order to study the dynamics of
  tracer particles, e.g.~stellar streams \citep{Lo11}. This technique
  continues to find use in dynamical studies of the Milky Way and
  other galaxies \citep{Sa20,Brooks2024}.  }%

\refedit{%
  However the original motivation for basis sets was to express the
  normal modes arising in the perturbation theory of self-gravitating
  systems \citep{Ka76,Sa91}. Each potential basis function
  is Fourier transformed
  $\Phi_n(\vec{r}) \to \hat{\Phi}_n^{\nu}(\vec{J})$ with respect to
  action-angle coordinates $(\vec{J},\vec{w})$ (here $\nu$ represents
  a vector of angular wavenumbers); then the matrix
  \begin{equation}
    M_{n m} = \sum_\nu \int d^{3}\vec{J} \: \hat{\Phi}_n^{\nu}(\vec{J}) \: \overline{\hat{\Phi}_{m}^{\nu}(\vec{J})} \: R_\nu(\vec{J})
  \end{equation}
  represents the self-consistent response of the system to the
  $\varrho_n$-density component of a perturbation ($R_\nu(\vec{J})$ is
  an operator encoding the microscopic behaviour of the particles in
  the model, e.g.~collisionless). Development of the method is ongoing
  \citep{Pe24,Tep2025}.%
}

\refedit{%
  In the context of computational chemistry (in particular
  Hartree-Fock theory, Kohn-Sham DFT, etc.), one of the central
  bottlenecks is the evaluation of the complex multi-electron
  interaction integrals which are required at each step of the
  iterative solution algorithm for Schrödinger's equation. In the LCAO
  approach the wavefunction is modelled as a linear combination of $N$
  atomic orbitals
  \begin{equation}
    \Psi_k = \sum_{j=1}^N c_j \psi_{j,k}, \qquad \psi_{j,k} = \psi_j(\vec{r} - \vec{a}_k).
  \end{equation}
  The interaction integrals
  $\langle \Psi_a \Psi_b, \Psi_c \Psi_d \rangle$ therefore involve a
  sum of terms of the form
  $\langle \psi_{1,a} \psi_{2,b}, \psi_{3,c} \psi_{4,d}\rangle$, which
  scales as $N^4$ in the number of orbitals (higher $N$ means a more
  accurate approximation of the wavefunction). Let us take a generic
  product of atomic orbitals $f = \psi_{1,a} \psi_{2,b}$ (with
  multipole moments $f_{lm}$) and attempt to evaluate the interaction
  integral \eqref{eq:self} using a multipole expansion of the Coulomb
  potential. We obtain the unwieldy nested indefinite integral
  \begin{equation}
    \langle f, f \rangle = \sum_{lm} \frac{4\pi}{2l+1} \int_0^\infty \overline{f_{lm}(r^\prime)}
    \left[ r^{\prime \: 1-l} \int_0^{r^\prime} r^{l+2} \: f_{lm}(r) \: dr + r^{\prime \: l+2} \int_{r^\prime}^\infty r^{1-l} \: f_{lm}(r) \: dr \right] dr^\prime.
  \end{equation}
  A similar situation obtains when using multipole-like expansions in
  other coordinate systems\footnote{\refedit{In particular, the
      diatomic Schrödinger equation separates in prolate spheroidal
      coordinates. Hence one may attempt to use the Neumann expansion
      to compute interaction integrals over (di-)atomic orbitals
      expressed in prolate spheroidal coordinates
      \citep{Mendl2012}.}}, so application of an appropriate Coulomb
  resolution is often more computationally efficient than numerical
  evaluation of these double integrals.%
}

\refedit{%
  Of course, this is of little concern if the chosen orbitals have a
  known closed-form interaction integral, as is the case with the
  well-known Gaussian orbital\footnote{\refedit{Closed-form solutions
      for the interaction integral are known for a small number of
      orbitals expressible in spherical polar coordinates -- for
      example the Gaussian and Slater-type orbitals -- but not for any
      in prolate spheroidal coordinates.}}
  $\psi_{j,k} = \expe^{-\gamma_j\lVert\vec{r} - \vec{a}_k\rVert^2}$.
  Nevertheless, sometimes such a large number of orbitals is required
  that the $N^4$ scaling of the overall interaction integral is itself
  problematic; then it is preferable to approximate each pair
  $\psi_{1,a} \psi_{2,b}$ with $n$ terms of an \textit{auxiliary basis}
  $\chi_k$\footnote{\refedit{In the literature this ubiquitous
      technique is referred to variously as `density fitting' or
      `resolution of the identity' (sometimes `RI' or `RI-J')
      \citep{Skylaris2000,Manby2001,Wirz2017}.}},
  \begin{equation}
    \varrho = \sum_{i,j=1}^N c_i c_j \psi_{i,a} \psi_{j,b} \approx \tilde\varrho = \sum_{k=1}^n d_k \chi_k,
  \end{equation}
  hence reducing the scaling to $N^2n^2$. Typically the coefficients
  $d_k$ are chosen such that the energy difference
  $\langle \varrho - \tilde\varrho, \varrho - \tilde\varrho \rangle$
  is minimised. Construction of such an energy-minimised auxiliary
  basis is therefore equivalent to Gram-Schmidt orthogonalisation of
  the auxiliary basis with respect to the energy inner product
  \eqref{eq:self}, an expensive step which is avoided if we use a
  basis that is already energy-orthogonal -- i.e.~a Coulomb
  resolution. While it is common in the literature to choose a
  non-orthogonal auxiliary basis with the same overall functional form
  as the underlying atomic orbitals (typically sums of Gaussians),
  this is not the only possibility.
  The Coulomb resolutions studied so far in the chemistry
  literature are those of \citet{Varganov2008} and \citet{Gill2009},
  both with ad-hoc functional forms originating from the Fourier
  transforms of Hermite or Laguerre polynomials respectively.
  Taking inspiration from the galactic dynamics approach, one could
  instead consider adapting the zeroth-order density function
  $\varrho_{0}$ exactly to an atomic orbital. One that is
  \textit{Gaussian-adapted}\footnote{\refedit{These Gaussian-adapted
      density basis functions are of the form
      $\rho_{nlm}(\vec{r}) = \expe^{-r^2} \! \times \!
      (\mathrm{polynomial \: in \:} r^2)\! \times \!
      Y_{lm}(\unitvector)$. See Sec.~\ref{sec:spherical_examples} of
      the present work for a similar example `adapted' to Slater-type
      orbitals.}} in this way was given in
  \citet[Sec.~4.1.3]{Lilley2023}, and the expansion coefficient in
  this basis for a radial Gaussian orbital (with different exponent)
  is
  \begin{equation}
  \langle \varrho_{n00}(\gamma r), \expe^{-\gamma_j r^2} \rangle \propto \frac{1}{\gamma_j \sqrt{\gamma + \gamma_j}} \left(\frac{\gamma - \gamma_j}{\gamma + \gamma_j}\right)^{n}.
  \end{equation}
  This is numerically well-behaved at arbitrarily high $n$ for any
  combination of $(\gamma, \gamma_j)$, and therefore likely compares
  favourably to any non-orthogonal auxiliary basis, or indeed any
  other Coulomb resolution.
}

\subsection{\refedit{Prior work}}

One way to phrase the search for \refedit{basis sets (or Coulomb resolutions)}
is to ask a slightly contrived question:~~\emph{How can we find a complete set of functions
  satisfying \eqref{eq:orth} and \eqref{eq:resolution}, such that
  \begin{enumerate}
  \item the summation in \eqref{eq:resolution} is `one-ranged',
    i.e.~not split or `two-ranged' with respect to the radial
    variable; and,
  \item the variables separate in a coordinate system in which the
    eigenfunctions of the Laplacian are unsuitable or unavailable?
  \end{enumerate}} The first requirement excludes the classical
multipole expansion.
Similarly, the compact expansions for rotational coordinate systems based on the
Heine formula \citep{Cohl1999} are excluded as they are two-ranged and
the variables are not separated when $z \neq 0$.

We identify three broad classes of approach: \textbf{(I)}
transformation of Poisson's equation to an eigenvalue problem, either
differential (Sturm-Liouville) or integral. Exceptionally (with a
special change of variable\footnote{The general form is
  $r \mapsto (r^{\alpha}-1)/(r^{\alpha}+1)$ as identified by
  \citet{Zh96}. Curiously this combines a stereographic transformation
  (familiar from momentum-space Hydrogen wavefunctions) with a varying
  exponent that arises from considering fractional Hankel transforms
  (see footnote \ref{fn:fractional}).}) this can result in a known
system of eigenfunctions, and hence some closed-form
results exist involving Gegenbauer polynomials: the basis sets of
\citet{CB73,HO92,Zh96} all use \refedit{essentially the same special}
change of variables -- but this exhausts the reasonable possibilities
\citep[Ch.~2]{lilley_thesis}. Therefore in general a discretisation of
the radial variable is required, which is the approach of
\citet{Bro98,We99}.

Alternatively \textbf{(II)} one may simply posit some
presumed-complete set of non-orthogonal functions, then orthogonalise
them numerically via the Gram-Schmidt process. This is the approach of
\citet{Sa91,Sa93,Bro98,Ea96,Ro96} -- the downsides being the
ill-conditioned nature of Gram-Schmidt orthogonalisation, and the
uncertain completeness of the resulting basis
functions\footnote{\refedit{In the (infinite-dimensional) Hilbert
    space of density functions of finite self-energy, a basis
    constructed through Gram-Schmidt orthogonalisation of an arbitrary
    collection of non-orthogonal elements is only guaranteed to span
    the same subspace as the original elements; without developing
    techniques such as those of Sec.~\ref{sec:basis}, it is difficult
    to prove that such a basis could span the whole space.}}.

Finally \textbf{(III)} one could hope for an invertible transform that
puts the problem in a more tractable form, ideally by transforming to
a space where more candidate orthogonal functions are known. These can
then be inverse-transformed back in the hope that the resulting
real-space orthogonal functions are satisfactory. There is a natural
option here, which is to use the eigenfunctions of the Laplacian
itself, i.e.~solutions to the Helmholtz equation
$\nabla^2 \psi_k = k^2 \psi_k$ (such $\psi_k$ are normally referred to
as the `wavefunctions' associated to a particular separable coordinate
system). As eigenfunctions of a self-adjoint operator, they
automatically obey an orthogonality relation
\begin{equation}
  \intddd \: \psi_k \: \overline{\psi_{k^\prime}} = N(k) \: \delta(k - k^\prime).
\end{equation}
Hence suitable basis functions can be constructed by choosing some
auxiliary functions $f_n(k)$ (not necessarily polynomials) that
satisfy their own orthogonality relation in $k$-space
\begin{equation}
  \int dk \: k^2 \: f_n(k) \: \overline{f_m(k)} = \delta_{nm},
\end{equation}
and then writing (neglecting the angular dependence)
\begin{equation}
  \Phi_n(\vec{r}) = \int dk \: f_n(k) \: \psi_k(\vec{r}), \qquad \varrho_n(\vec{r}) = 4\pi \int dk \: k^2 \: f_n(k) \: \psi_k(\vec{r}).
\end{equation}
One then hopes that $\varrho_0$ describes the zeroth-order of the
particular system one is interested in studying. In fact, due to some
good fortune with auxiliary functions, this method has been used to
find several closed-form basis sets, especially in spherical (or
circular) polar coordinates where the wavefunctions are proportional
to (spherical) Bessel functions, for which a multitude of appropriate
integrals (Hankel transforms) are readily available\footnote{The case
  of infinitesimally thin discs is formally similar to spherical polar
  coordinates, due to the wavefunctions derived by
  \citet{Too63}.}. This was the approach of
\citet{Ra09,LSEE,LSE,Varganov2008} in the spherical polar case; and of
\citet{CB72,Qi93} in the case of infinitesimally thin discs in
circular polar coordinates. Additionally if one is interested in
systems of finite extent, then no integral transform is required and
finite combinations of wavefunctions can be used,
e.g.~\citet{Tre76,Po81}.

However, as soon as we look further afield than spherical polar
coordinates, we start to run into problems. In the spheroidal and
ellipsoidal coordinate systems the corresponding wavefunctions are
highly non-trivial, with no corresponding integral transforms known
\citep{Arscott1981}, and they are rarely implemented in numerical
libraries. Worse, in an $R$-separable coordinate system the Helmholtz
equation does not separate, and so the wavefunctions do not even
exist! Therefore in order to make use of these coordinate systems that
may be better suited to the geometry of the system under study, we are
forced to find a method that does not directly rely on eigenfunctions
of the Laplacian. The subject of this paper is essentially a way to
generalise method \textbf{(III)} to various other separable coordinate
systems in a way that avoids relying on the eigenfunctions of the
Laplacian.  As we shall describe in Sec.~\ref{sec:operators}, our
solution is to diagonalise an alternative set of operators, chosen to
be in some way compatible with the Laplacian. Then we choose a
potential-density pair that defines the zeroth-order of the new basis
set; and we define the higher-order terms of the basis by requiring
that one of the operators is tri-diagonal in the new basis.

The first realisation of this method was in spherical polar
coordinates \citep{Lilley2023}. We replaced the Hankel transform with
a Mellin transform, already known for solving Laplace's equation in
wedge-shaped domains \citep{Sneddon1972}. We showed that not only do
essentially all the extant closed-form basis functions (types
\textbf{(I)} or \textbf{(III)}) arise naturally with our new method,
but also that we attain a systematic semi-numerical way of producing
basis functions with an almost arbitrary choice of (differentiable)
zeroth-order function. This freedom was not available in the Hankel
transform method except through luck in choosing appropriate auxiliary
functions\footnote{We also impose that the auxiliary functions must be
  polynomials (a guarantee which simplifies the construction of the
  resulting basis functions), which is not in general true of the
  Hankel transform method.}\footnote{\label{fn:fractional}As early
  evidence that the Hankel transform might not exhaust the
  possibilities for forming closed-form basis sets, it was already
  observed in \citet[Appendix, Eq.~A34]{Po81} that an additional
  degree of freedom arises by considering fractional-order Hankel
  transforms. This allows the power-law behaviour of the resulting
  basis sets to be adjusted, which was later put to use
  \citep{LSEE,LSE}. But these fractional-order Hankel transforms do
  not arise naturally from the general consideration of solutions to
  the Helmholtz equation in separable coordinates. In contrast, in the
  present operator formalism the fractional parameter arises simply as
  a multiplier to the argument of the index-raising polynomial
  \citep[Sec.~4.1.2]{Lilley2023}.}. Now we demonstrate that this
method generalises to several other separable coordinate systems, at
the expense of more complicated operators and integral transforms.

\section{Operators from separable coordinates}\label{sec:operators}

In the following we assume that all mass or charge densities $\varrho$
are infinitely smooth everywhere and vanish at infinity, although
other boundary conditions are possible\footnote{The other geometry
  that has been studied is the `thin disc' geometry where
  $\varrho = 0$ for $|z|>0$ has been considered in
  \citet{Ka76,Qi93,Lilley2023}, among others. See discussion in
  Sec.~\ref{sec:discussion}.}.

According to \citet[Ch.~3]{Miller1984}, separable coordinate systems
for both Laplace's and Helmholtz's equations can be classified by
commuting pairs of operators $(S_1, S_2)$ where both operators are
symmetric second-order generalised symmetries of the Laplacian. The
separable solutions are then simultaneous eigenfunctions of $S_1$ and
$S_2$.

There are two types of separable coordinate systems: first, systems
which are regularly separable for both Laplace's and Helmholtz's
equations, where the operators $S_j$ involve only translations and
rotations (which commute with the Laplacian); and second, systems
which are $R$-separable for Laplace's equation but \emph{not}
separable for Helmholtz's equation, where the $S_j$ involve the full
set of quadratic conformal symmetries (including the dilatation and
special conformal transformations). In addition, if all the coordinate
surfaces possess a rotational symmetry around one axis the coordinate
system is referred to as `rotational'; if there is a translational
symmetry along one axis then it is `cylindrical'. Conventionally this
symmetry axis is always taken to be the $z$ axis and the corresponding
symmetry operator is $S_2$; so in the former case we have
$S_2 = \jop_3$ (which generates $z$-rotations), and in the latter
$S_2 = \pop_3$ (which generates $z$-translations). Here we restrict
ourselves to the case of rotational or cylindrical coordinate systems,
either simply or $R$-separable.

Any generalised Laplacian symmetry operator (listed in
App.~\ref{sec:algebra}) maps the space of solutions of Laplace's
equation to itself, and in particular the operator $T$ represents the
same map as $T + f\Lap$, for an arbitrary function $f$. Hence when
studying solutions to Laplace's equation, the operators $T$ and
$T + f\Lap$ are taken to be equivalent, and any commuting pair of
generalised symmetry operators are equivalent to the pair $(S_1,S_2)$
that define some separable coordinate system.

However, as we are concerned with constructing a basis to solve
Poisson's equation, we no longer have equivalence between $T$ and
$T + f\Lap$. Therefore we can seek commuting triplets of inequivalent
operators $(T_1, T_2, T_3)$ and use their mutual eigenfunctions as a
basis for solutions (see App.~\ref{sec:Sj_Tj_relationship} for some
discussion of the relationship between $T_j$ and $S_j$).

\refedit{Let $T^*$ be the adjoint of $T$ with respect to the standard
$L^2(\mathbb{R}^3)$ inner product,
\begin{equation}
  (T f, g) = (f, T^* g), \qquad\qquad (f,g) = \intddd \: f(\vec{r}) \: \overline{g(\vec{r})}.
\end{equation}
We require only those operators which
\textit{quasi-commute}\footnote{\refedit{These could also be called
  \emph{almost-commuting} operators. They are also reminiscent of the
  \textit{intertwining} operators of \citet{Anderson1990}, although the
  scenario is reversed: they seek an intertwining operator $D$ such
  that $LD = D L^\prime$ for two given operators $L$ and $L^\prime$;
  for us the intertwining operator is fixed to be the Laplacian, and
  it is the other two operators that must be found, subject to certain
  conditions.}}  with the Laplacian, i.e.~satisfying
\begin{equation}\label{eq:quasi}
 \Lap T^* = T \Lap.
\end{equation}
This is accomplished by constructing each $T$ out of symmetric
quadratic Laplacian symmetries, including terms of the form $f \Lap = (\Lap f)^*$.
The quasi-commutation property suffices to make $T$ symmetric with
respect to the energy inner product \eqref{eq:self},
\begin{equation}
\langle T f, g \rangle = \langle f, T g \rangle.
\end{equation}}
We will now construct (non-uniquely) a commuting set of operators
$T_j$ associated with a rotational $R$-separable coordinate system. We
mostly use the standard notation \mf{Ch.~5}, with some slight
adjustments.

Given $(q_1,q_2,q_3)$, a rotational $R$-separable coordinate system,
we label $q_1$ the `radial' coordinate, $q_3$ the `azimuthal'
coordinate, and generically $q_2$ as another `angular'
coordinate. Associated with such a coordinate system are ten
quantities: three scale factors $h_n$, three factors $f_n$ (that
appear in the Stäckel determinant $S$), three Stäckel matrix minors
$M_n = S/h_n^2$, and the modulation factor $R$. Because we are only
considering rotational or cylindrical coordinate systems, we can
assume that these quantities depend on the coordinates as
\begin{equation}
  h_n(q_1,q_2), f_n(q_n), M_1(q_2), M_2(q_1), R(q_1,q_2),
\end{equation}
and $f_3 = -1$. If each variable is re-parameterised according to
$q_n \mapsto x_n(q_n)$, we have the transformation rules
$h_n \mapsto h_n(d q_n / d x_n)$ and
$f_n \mapsto f_n/(d q_n / d x_n)$; this lets us rewrite Morse \&
Feshbach's standardised variables $q_n$ in terms of more familiar
variables. For example, in the case of spherical polar coordinates, this means
transforming
$(q_1, q_2, q_3) = (r, \cos\vartheta, \cos\varphi) \mapsto (r,
\vartheta, \varphi)$.

In the given coordinate system the Laplacian becomes
\begin{equation}
\Lap = \sum_n L_n, \qquad L_n = h_n^{-2}\left( \partial_n \log{\left(R^2 f_n\right)} \partial_n + \partial_n^2 \right),
\end{equation}
so that $L_n$ is the term that contains all the derivatives in
$\partial_n \equiv \partial/\partial {q_n}$. Now we introduce a new
function
\begin{equation}
g(q_1, q_2) = \frac{f_1 f_2 h_1 h_2 R^2}{f_3 h_3},
\end{equation}
and two more functions $v_n(q_n)$ which are the terms in $g/h_3^2$
dependent on each coordinate,
\begin{equation}\label{eq:v_def}
  v_1(q_1) + v_2(q_2) = \frac{g}{h_3^2}.
\end{equation}
Then we define the trio of operators $T_j^*$,
\begin{align}\label{eq:Tj_def}
  T_1^* &= -g L_1 + v_1 T_3^2 + u, \\ \nonumber
  T_2^* &= -g L_2 + v_2 T_3^2 - u, \\ \nonumber
  T_3^* &= T_3 = \sqrt{-h_3^2 L_3} \quad (= \jop_3 \:\:\text{or}\:\: \pop_3),
\end{align}
and choose $u(q_1,q_2)$ such that $T_1^*$ and $T_2^*$ commute; one possible
choice (see App.~\ref{sec:operator_deriv}) is
\begin{equation}\label{eq:u_def}
  u = \frac{1}{2} \partial_1^2 \log{R^2} + \frac{1}{4} \left( \partial_1\log{\left(R^2 f_1\right)} \right)^2.
\end{equation}
These operators satisfy
\begin{equation}\label{eq:t1_t2_lap}
T_1^* + T_2^* = -g\Lap.
\end{equation}
However, if it turns out that $T_1^*$ is a perfect square in the
operator sense (perhaps after adding a constant $c$), then we can
instead define
\begin{equation}
  T_1^* = \sqrt{-g L_1 + v_1 T_3^2 + u + c},
\end{equation}
and replace $T_1^*$ with $T_1^{*2} - c$ in the preceding relation
\eqref{eq:t1_t2_lap} (and similarly for $T_2^*$).

As the three operators mutually commute we can find their joint
eigenfunctions: the \textit{eigen-potentials} $\phi_{\eivec}$ obeying
\footnote{\refedit{The eigen-potentials are related to the
    corresponding solutions to Laplace's equation, but they depend on
    all three separation constants $\eivec_j$ (see discussion in
    App.~\ref{sec:Sj_Tj_relationship}).}}
\begin{equation}\label{eq:eipot}
  T_j^* \phi_{\eivec} = \eivec_j \phi_{\eivec}, \qquad \eivec = (\eivec_1, \eivec_2, \eivec_3)
\end{equation}
and the \textit{eigen-densities} $\Psi_{\eivec}$ satisfying
\begin{equation}\label{eq:eidens}
  T_j \Psi_{\eivec} = \eivec_j \Psi_{\eivec}, \qquad  \Lap \phi_{\eivec} = 4 \pi \Psi_{\eivec}.
\end{equation}
Each $T_j$ quasi-commutes with the Laplacian, so the eigen-potentials
and eigen-densities are related by
\begin{equation}\label{eq:eigen_potential_density_relation}
  \phi_\eivec = \frac{-4 \pi g}{\eivec_1 + \eivec_2} \Psi_\eivec,
\end{equation}
or introducing $c$ into the denominator as necessary. The
quasi-commutation property implies the $T_j$ are symmetric in the
energy inner product \eqref{eq:self}, and so their eigenfunctions
$\Psi_\eivec$ are necessarily orthogonal in the same inner product, so
that
\begin{equation}\label{eq:psi_orthog}
  \langle \Psi_{\eivec}, \Psi_{\eivec^\prime} \rangle = F_\eivec \: \delta_{\eivec \eivec^\prime},
\end{equation}
with \refedit{$F_\eivec$ a real normalisation} factor that must be determined.

\section{Basis sets from operators}\label{sec:basis}

\refedit{%
  Now we use the operators $T_j$ defined in Sec.~\ref{sec:operators}
  to construct both a basis set \eqref{eq:orth} and a Coulomb
  resolution \eqref{eq:resolution} associated with a given separable
  coordinate system $(q_1,q_2,q_3)$. The azimuthal angular variable
  $q_3$ is restricted to $(0, 2\pi)$, and so its eigenvalue
  $m \equiv \sqrt{\eivec_3}$ takes integer values. We assume that the
  radial variable $q_1$ has infinite range, and so its corresponding
  eigenvalue $s \equiv \eivec_1$ or $\sqrt{\eivec_1}$ is
  continuously-valued on the spectrum $\Omega = (0, \infty)$ or
  $(-\infty, \infty)$. We also assume here that the remaining variable
  $q_2$ is `angular', which is to say it is restricted in range; if it
  is not already naturally restricted, then we must take periodic
  boundary conditions with respect to $q_2$, so that
  $l \equiv \sqrt{\eivec_2}$ is also restricted to integer
  values. Hence the eigenvalue $\lambda$ from Sec.~\ref{sec:operators}
  is now written $slm$. We use $\sum_{lm}$ or $\bigoplus_{lm}$ to
  indicate whichever sum over all the angular eigenvalues makes sense
  in a given coordinate system (for example,
  $\sum_{lm} \equiv \sum_{l=0}^\infty \sum_{m=-l}^{l}$ in spherical
  polar coordinates).%
}

\refedit{%
  For the following exposition we take
  \citet[Ch.~VII--VIII]{ReedSimon1} as our source. Let $\mathcal{H}$
  be the Hilbert space of density functions with inner product
  \eqref{eq:self} and finite self-energy\footnote{\refedit{Requiring
      that only the total self-energy be finite means that some
      density functions -- well-known from galactic modelling -- are
      included, that might be excluded under other natural-seeming
      criteria. For example, the \citet{NFW} density
      ($\varrho \propto r^{-1}(1+r)^{-2}$) has infinite total mass;
      and \citet{De93} models
      ($\varrho \propto r^{-\gamma}(1+r)^{\gamma-4}$) with
      $\gamma \geq 3/2$ are not square-integrable.}},
\begin{equation}
  \mathcal{H} = \{ \varrho \: | \: \langle \varrho, \varrho \rangle = \lVert \varrho \rVert^2 < \infty \},
\end{equation}
and given eigenfunctions $\Psi_{slm}$ defined as in \eqref{eq:eidens},
let $P_{lm}$ be the orthogonal projection with respect to the
$(l,m)$-th angular eigenfunction,
\begin{equation}
  P_{lm}(\varrho) = \int_\Omega ds \: \frac{\Psi_{slm} \: \langle \varrho, \Psi_{slm} \rangle}{F_{slm}}, \qquad P_{lm}(P_{\lambda\mu}(\varrho)) = \delta_{l\lambda} \delta_{m\mu} P_{lm}(\varrho).
\end{equation}
This lets us decompose $\mathcal{H}$ into a direct sum of subspaces,
one for each combination of angular eigenvalues,
\begin{equation}
  \mathcal{H} = \bigoplus_{lm} \mathcal{H}_{lm}, \qquad \mathcal{H}_{lm} = P_{lm}(\mathcal{H}).
\end{equation}
Now we assume that each $\mathcal{H}_{lm}$ admits a cyclic vector with
respect to the operator $T_1$, and take $\{ \varrho_{0lm} \}$ to be a
particular choice of these vectors (by the spectral theorem, the lack of repeated
eigenvalue $s$ on each $\mathcal{H}_{lm}$ is equivalent to the
existence of at least one cyclic vector). An arbitrary
$\varrho \in \mathcal{H}_{lm}$ is therefore equal to
$f(T_1)\varrho_{0lm}$ for some function $f(s)$ that can be
approximated as closely as desired by a polynomial (with respect to a
certain weight $\omega_{lm}(s)$). In fact, $f(s)$ is given by the
unitary integral transform
$U_{lm} : \mathcal{H}_{lm} \to L^2(\Omega,\omega_{lm})$, where
\begin{equation}\label{eq:Ulm_transform}
  U_{lm}(\varrho)(s) = \frac{\langle \varrho, \Psi_{slm}\rangle}{\langle \varrho_{0lm}, \Psi_{slm}\rangle},
\end{equation}
and the corresponding measure is
\begin{equation}\label{eq:subspace_integration}
  \lVert U_{lm}(\varrho) \rVert^2_{\omega_{lm}} = \int_\Omega ds \: |U_{lm}(\varrho)(s)|^2 \: \omega_{lm}(s), \qquad \omega_{lm}(s) = \frac{\langle \Psi_{slm}, \varrho_{0lm} \rangle}{F_{slm}}.
\end{equation}
Now extend this to a transform on all of $\mathcal{H}$, namely
$U : \mathcal{H} \to L^2(\Omega,\omega)$, where
\begin{equation}
  U = \bigoplus_{lm} U_{lm} \circ P_{lm},
\end{equation}
with corresponding measure
\begin{equation}
  \lVert U(\varrho) \rVert^2_{\omega} = \sum_{lm} \lVert U_{lm}(P_{lm}(\varrho)) \rVert^2_{\omega_{lm}} = \lVert \varrho \rVert^2.
\end{equation}
So $U$ preserves the self-energy of $\varrho$, and operating by $T_1$
becomes multiplication by $s$ under $U$. In practice however $T_1$ is
unbounded on $\mathcal{H}$\footnote{\refedit{This can be shown
    directly by taking one of the basis sets discussed in
    Sec.~\ref{sec:coords} that has an explicit formula involving known
    classical orthogonal polynomials, and observing that
    $\lVert T_1 \varrho_{nlm} \rVert^2/\lVert \varrho_{nlm} \rVert^2$
    grows at least linearly with $n$; there is no upper bound to $n$,
    therefore $T_1$ is unbounded.}}, so for the above to work we need
$T_1$ to be essentially self-adjoint. We do not attempt to prove this
in general, but instead reason backwards from the knowledge of $U$ in
several important cases: in spherical polar coordinates $U$ is the
Mellin transform (Sec.~\ref{sec:sph_pol}); in cylindrical polar
coordinates the Kontorovich-Lebedev transform (Sec.~\ref{sec:cyl});
and in prolate spheroidal coordinates the Mehler-Fock transform
(Sec.~\ref{sec:pro_sph}). With a suitable normalisation these are all
well-defined unitary integral transforms \citep{Sneddon1972}, so we
conclude that $T_1$ is essentially self-adjoint and so the spectral
theorem applies in these cases (for bispherical and toroidal
coordinates we can use the same reasoning, but for the remaining
coordinate systems of Sec.~\ref{sec:coords} we merely argue
heuristically).
}

\refedit{%
  Next we construct an orthogonal basis for each $\mathcal{H}_{lm}$ by
  first constructing the (unique up to an overall factor) basis of
  orthogonal polynomials with respect
  to the weight function $\omega_{lm}$ on $\Omega$. The measure
  $\lVert \cdot \rVert_{\omega_{lm}}$ \eqref{eq:subspace_integration}
  becomes an inner product on
  polynomials by polarisation. If the moments $\mu_{nlm}$ are bounded,
  \begin{equation}\label{eq:moments}
    \mu_{nlm} = \lVert s^{n} \rVert_{\omega_{lm}} \leq \infty,
  \end{equation}
  and the Hankel determinants $\Delta_{nlm}$ are non-vanishing,
  \begin{equation}\label{eq:hankel_determinants}
    \Delta_{nlm} = \left| \begin{matrix} \mu_{0lm} & \mu_{1lm} & \ldots & \mu_{n-1,lm} \\
      \mu_{1lm} & \mu_{2lm} & \ldots & \mu_{nlm} \\
      \vdots & \vdots &  & \vdots \\
      \mu_{n-1,lm} & \mu_{nlm} & \ldots & \mu_{2n-2,lm} 
    \end{matrix}\right| > 0,
  \end{equation}
  then we can construct the orthogonal polynomials $p_{nlm}$ using
  standard methods \citep{Gautschi1985,Gautschi1990}. We call
  $p_{nlm}$ the \textit{index-raising polynomials} associated with the
  particular $\varrho_{0lm}$, and they satisfy
\begin{equation}
  \int_\Omega ds \: p_{nlm}(s) \: \overline{p_{\nu l m}(s)} \: \omega_{lm}(s) = \delta_{n\nu} N_{nlm}.
\end{equation}
Then, transferring this basis back to $\mathcal{H}$, we
have the \textit{density basis functions} (which tridiagonalise $T_1$)
\begin{equation}
  \varrho_{nlm} = U^{-1}(p_{nlm}) = p_{nlm}(T_1) \varrho_{0lm}, \qquad \langle \varrho_{nlm}, \varrho_{\nu\lambda\mu} \rangle = \delta_{n\nu} \delta_{l\lambda} \delta_{m\mu} N_{nlm},
\end{equation}
and so any $\varrho \in \mathcal{H}$ has the expansion
\begin{equation}
  \varrho = \sum_{lm} \sum_{n=0}^\infty c_{nlm} \: \varrho_{nlm}, \qquad c_{nlm} = \langle \varrho, \varrho_{nlm} \rangle/N_{nlm}, \qquad \lVert \varrho \rVert^2 = \sum_{lm} \sum_{n=0}^\infty |c_{nlm}|^2 N_{nlm}.
\end{equation}
If the polynomials $p_{nlm}$ are complete and dense in
$L^2(\Omega,\omega_{lm})$, then the resulting expansions converge in
the norm $\lVert \cdot \rVert_{\omega_{lm}}$. So by the unicity of
$U$, approximations by $\varrho_{nlm}$ of elements of $\mathcal{H}$
are convergent in the self-energy norm; this avoids the need to
directly check whether each $\varrho_{0lm}$ is a cyclic vector for
$\mathcal{H}_{lm}$. Of course, the conditions
\eqref{eq:moments}--\eqref{eq:hankel_determinants} are rather
technical, but for a given choice of $\varrho_{0lm}$ they can be
checked numerically up to a certain $n$. Informally, `reasonable'
choices of $\varrho_{0lm}$ are valid -- those which have infinite
support, are monotonically decreasing and possess derivatives of all
orders\footnote{Formally we must also have
  $\sum_{lm} \lVert \varrho_{0lm} \rVert^2 < \infty$, but this can be
  achieved by just changing the $(l,m)$-dependence of the
  normalisation.}.%
}

\refedit{%
  Because $T_1$ is quasi-commuting, given the solution to
  $\Lap \Phi_{0lm} = 4 \pi \varrho_{0lm}$\footnote{One possible way to
    find $\Phi_{0lm}$ is to invert the representation
    $\Phi_{0lm} = \sum_{lm} \int_\Omega ds \: \phi_{slm} \: \langle
    \varrho_{0lm}, \Psi_{slm} \rangle$; then by
    \eqref{eq:eigen_potential_density_relation} this just means evaluating
    $U^{-1}\!\left(\frac{g}{s^2 + l^2}U(\varrho_{0lm})\right)$ (of
    course this is not necessarily the easiest way to solve Poisson's equation).}, we can
  solve Poisson's equation for each $\varrho_{nlm}$ to find the
  higher-order \textit{potential basis functions}
\begin{equation}
  \Phi_{nlm} = p_{nlm}(T_1^*) \Phi_{0lm},
\end{equation}
which satisfy $\Lap \Phi_{nlm} = 4 \pi \varrho_{nlm}$ (by
quasi-commuting each $T_1$ past the Laplacian). We therefore have the
`resolution of the identity'
\begin{equation}
  \delta^3(\vec{r} - \vec{r}^\prime) = -\sum_{lm} \sum_{n=0}^\infty \varrho_{nlm}(\vec{r}) \: \overline{\Phi_{nlm}(\vec{r}^\prime)}/N_{nlm},
\end{equation}
and similarly the `resolution of the Coulomb operator'
\begin{equation}
\frac{4\pi}{\lVert \vec{r} - \vec{r}^\prime \rVert} = \sum_{lm} \sum_{n=0}^\infty \Phi_{nlm}(\vec{r}) \: \overline{\Phi_{nlm}(\vec{r}^\prime)} / N_{nlm}.
\end{equation}
We can always take $p_{nlm}$ to be monic,
obeying the recurrence relation
\begin{equation}\label{eq:pnlm_recurrence}
  p_{n+1,lm}(s) = (s-\alpha_{nlm})\:p_{nlm}(s) - \beta_{nlm} \: p_{n-1,lm}(s),
\end{equation}
so the basis functions $\varrho_{nlm}$ and $\Phi_{nlm}$ obey similar
(differential-recurrence) relations, obtained by substituting $T_1$ or
$T_1^*$ for the variable $s$,
\begin{align}\label{eq:diff_recur}
  {\Phi}_{n+1,lm} &= (T_1^* - \alpha_{nlm}) {\Phi}_{nlm} - \beta_{nlm} \: {\Phi}_{n-1,lm} \\ \nonumber
  {\varrho}_{n+1,lm} &= (T_1 - \alpha_{nlm}) {\varrho}_{nlm} - \beta_{nlm} \: {\varrho}_{n-1,lm}.
\end{align}
If nothing more is known about $\alpha_{nlm}$ and $\beta_{nlm}$ then
each higher-order basis function can be produced from the two
preceding functions, either laboriously by manual calculation or
numerically via a technique such as \textit{automatic
  differentiation}. However, in the case where $p_{nlm}$ is a
classical orthogonal polynomial, the recurrence coefficients become
simple, and we typically find that more compact closed-form
expressions for the basis functions exist in terms of known classical
polynomials and special functions, which come equipped with new
recurrence relations whose numerical behaviour is much easier to study
than the differential-recurrence relations \eqref{eq:diff_recur}.
  Construction of the basis set in this way is equivalent to
  Gram-Schmidt orthogonalisation of the `monomials' $T_1^n \varrho$ (a
  type \textbf{(II)} basis set in the nomenclature of
  Sec.~\ref{sec:intro}).%
}

\refedit{%
  Analysis of the stability of
  \eqref{eq:diff_recur} is complicated by its differential nature.
  Generating the underlying polynomials (i.e.~the coefficients
  $\alpha_{nlm}$ and $\beta_{nlm}$, and the normalisation constant
  $N_{nlm}$) is straightforward using the techniques of
  \citet{Gautschi1990} (for purely numerical construction we found
  that the \textit{discretised Stieltjes procedure} wins out; see
  discussion in \citet[Sec.~5]{Lilley2023}). However the repeated
  differentiations (applications of $T_1$ or $T_1^*$) can be
  problematic, and depend on the specific functional forms at hand. If
  these differentiations are handled efficiently, then by
  invoking \eqref{eq:diff_recur} we are conceptually avoiding the
  $k-1$ re-projections that occur at step $k$ of the Gram-Schmidt
  process, so we would expect a more stable numerical performance.
}

\refedit{%
  It is sometimes possible to directly verify the Coulomb
  resolution itself. Starting with the completeness relation for the
  polynomials $p_{nlm}$,
\begin{equation}\label{eq:poly_complete}
\sum_{n=0}^\infty \frac{p_{nlm}(s) \: \overline{p_{nlm}(t)}}{N_{nlm}} = \frac{\delta(s - t)}{\omega_{lm}(s)},
\end{equation}
we have
\begin{equation}\label{eq:completeness_relation}
  \sum_{n=0}^\infty \sum_{lm} \frac{\Phi_{nlm}(\vec{r}) \: \overline{\Phi_{nlm}(\vec{r}^\prime)} }{ N_{nlm} }
  = \sum_{lm} \int_\Omega ds \frac{\phi_{slm}(\vec{r}) \: \overline{\phi_{slm}(\vec{r}^\prime)}}{F_{slm}}.
\end{equation}
Then in three cases we can find an explicit solution to the
$s$-integral, and match the resulting sum to a familiar multipole-like
formula: for spherical polar coordinates this is
\eqref{eq:sph_complete}, for cylindrical polar
\eqref{eq:cyl_complete}, and for prolate spheroidal
\eqref{eq:pro_complete}.%
}

\section{Coordinate systems}\label{sec:coords}

For each coordinate system we give both a reference to
\cite[Ch.~5]{MF} and its number in the classification established by
\cite[Ch.~3]{Miller1984}.

We included the cylindrical coordinate systems because the formalism
developed above basically works with minimal modification\footnote{For
  all cylindrical systems, at minimum the $z$ coordinate (and possibly
  other coordinates) must be restricted in range in order to
  produce a viable basis set formalism.}, although many of them are
less tractable (the least are relegated to
App.~\ref{sec:other_cylindrical}). We omit conical coordinates as they
have the same radial coordinate as spherical polar coordinates, so the
basis sets would only differ in the angular part. The Laplacian
symmetry operators (in curly letters) are defined in
App.~\ref{sec:algebra}; \refedit{their symmetric products are denoted
$\{\mathcal{A},\mathcal{B}\} = \mathcal{A}\mathcal{B} +
\mathcal{B}\mathcal{A}$}.
The modulation factor $R$ is relevant ($R \neq 1$) only for the
$R$-separable coordinates (bispherical and toroidal).
\refedit{Other notation: $(a)_n$ indicates the Pochhammer symbol,
  ${}_{p}{F}_{q}$ the generalised hypergeometric function, and
  $P_n^{(a)}(x)$ the Legendre function. Any other special functions will be
  noted as they appear.}

In general all these coordinate systems have a counterpart for
densities restricted to the plane $z=0$, which we summarise in
Sec.~\ref{sec:discussion} but do not discuss in detail here.

\subsection[Spherical polar]{Spherical polar \coordref{V}{5}}\label{sec:sph_pol}
\begin{flalign*}
\text{\textit{Coords:}} \quad & q_1 = r, \quad q_2 = \vartheta, \quad q_3 = \varphi && \\
                              & x = r\sin{\vartheta}\cos{\varphi}, \quad y = r\sin{\vartheta}\sin{\varphi}, \quad z = r\cos{\vartheta} && \\
                              & h_1 = 1, \quad h_2 = -r, \quad h_3 = -r \sin\vartheta, \quad f_1 = r^2, \quad f_2 = -\sin\vartheta, \quad g = r^2 && \\
\text{\textit{Operators:}} \quad & L_1 = 2 r^{-1} \partial_r + \partial_r^2, \quad L_2 = r^{-2}\left(\cot\vartheta \partial_\vartheta + \partial_\vartheta^2\right), \\
                              & T_1^* = \dop^* = \sqrt{-gL_1 - 1/4} = \imagi\left( r \partial_r + 1/2\right), \quad T_2 = \jop^2, \quad T_3 = \jop_3
\end{flalign*}
This is the simplest example of a first-order $T_1$, which can be
found practically by inspection due to its trivial coordinate
dependence. The identity \eqref{eq:t1_t2_lap} now reads
\begin{equation}\label{eq:sph_pol_lap_id}
\dop^{*2} + \jop^2 + 1/4 = -r^2\Lap.
\end{equation}
The eigen-potentials and -densities are proportional to spherical
harmonics\footnote{To align with conventional spherical harmonics we
  use $l+1/2$ as the second eigenvalue.}
\begin{align}
  \phi_{slm}(r,\vartheta,\varphi) &= \frac{-4 \pi \: r^{\imagi s - 1/2} \: P_{l}^{(m)}\!\left(\cos\vartheta\right) \: \expe^{\imagi m \varphi}}{s^2 + (l+1/2)^2}, \\ \nonumber
  \Psi_{slm}(r,\vartheta,\varphi) &= r^{\imagi s - 5/2} \: P_{l}^{(m)}\!\left(\cos\vartheta\right) \: \expe^{\imagi m \varphi},
\end{align}
and they satisfy the eigenvalue equations
\begin{equation}
  \left(\substack{\vphantom{s}T_1\\\vphantom{l(l+1)}T_2\\\vphantom{m}T_3}\right) \Psi_{slm} = \left(\substack{\vphantom{T_1}s\\\vphantom{T_2}l(l+1)\\\vphantom{T_3}m}\right) \Psi_{slm},
  \qquad\qquad \left(\substack{\vphantom{s}T_1^*\\\vphantom{l(l+1)}T_2^*\\\vphantom{m}T_3^*}\right) \phi_{slm} = \left(\substack{\vphantom{T_1^*}s\\\vphantom{T_2^*}l(l+1)\\\vphantom{T_3^*}m}\right) \phi_{slm}.
\end{equation}
The orthogonality relation
is\footnote{$J_{lm} = 2 \: (l+m)!/((2l+1) \: (l-m)!)$ is the
  normalisation constant for the spherical harmonics.}
\begin{equation}
\langle \Psi_{slm}, \Psi_{t \lambda \mu} \rangle = \frac{8 \pi^2 J_{lm}}{s^2 + (l + 1/2)^2} \delta_{m \mu} \delta_{l \lambda} \delta(s - t),
\end{equation}
so the expansion coefficient of $f$ in the basis of $\Psi_{slm}$
(the transform \eqref{eq:Ulm_transform}) is a Mellin transform in
$r$ and a spherical harmonic expansion in the angular variables.

The formalism for basis sets in spherical polar coordinates is
described in \citet{Lilley2023}; the conventional notation is
\begin{align}
  \Phi_{nlm}(r,\vartheta,\varphi) &= \Phi_{nl}(r) \: Y_{lm}(\vartheta,\varphi), \qquad \Phi_{nl} = p_{nl}(\dop^*) \Phi_{0l}, \\ \nonumber
  \varrho_{nlm}(r,\vartheta,\varphi) &= \varrho_{nl}(r) \: Y_{lm}(\vartheta,\varphi), \qquad \varrho_{nl} = p_{nl}(\dop) \varrho_{0l}, \\ \nonumber
  Y_{lm}(\vartheta,\varphi) &= P_{l}^{(m)}\!\left(\cos\vartheta\right) \: \expe^{\imagi m \varphi}/\sqrt{J_{lm}}.
\end{align}
We can verify the Coulomb resolution for such a basis set (following
\eqref{eq:poly_complete}--\eqref{eq:completeness_relation}),
\begin{align}\label{eq:sph_complete}
  \sum_{lm} \sum_{n=0}^\infty \frac{\Phi_{nlm}(\vec{r}) \: \overline{\Phi_{nlm}(\vec{r}^\prime)}}{N_{nl}}
  &= \sum_{lm} Y_{lm}(\unitvector) \: \overline{Y_{lm}(\unitvector^\prime)} \: \frac{2}{\sqrt{r r^\prime}} \int_{-\infty}^{\infty} ds \: \frac{\left(r^\prime/r\right)^{\imagi s}}{s^2 + (l+1/2)^2} \\ \nonumber
  &= \sum_{lm} Y_{lm}(\unitvector) \: \overline{Y_{lm}(\unitvector^\prime)} \: \frac{4 \pi \:  \expe^{-(l+1/2)|\log{(r^\prime/r)| }}}{(2l+1) \: \sqrt{r r^\prime}} \\ \nonumber
  &= \sum_{lm} Y_{lm}(\unitvector) \: \overline{Y_{lm}(\unitvector^\prime)} \: \frac{4 \pi}{2l+1} \times \begin{cases} r^l (r^\prime)^{-l-1}, & r < r^\prime \\ r^{-l-1} (r^\prime)^l, & r > r^\prime \end{cases},
\end{align}
where in the penultimate line we evaluated a Yukawa-type Fourier
integral; thus we recover the standard multipole expansion for
$1/\lVert \vec{r} - \vec{r}^\prime\rVert$.

Examples of spherical polar basis sets where $p_{nl}$ is a classical
orthogonal polynomial are: \citet{CB73} and the related developments
in \citet{HO92,Zh96,Ra09,LSEE}; and the generalisation of all these in
\citet{LSE} -- all having density functions of the `double-power law'
form familiar in galaxy modelling. Additionally some basis sets with
exponential fall-off in density have been found \citep[Sec.~4.1.3]{Lilley2023}.
In all these cases $p_{nl}$ is proportional to a Meixner-Pollaczek or
continuous Hahn polynomial, and compact formulas and recursion
relations for $\Phi_{nlm}$ and $\varrho_{nlm}$ exist which can be used
in place of \eqref{eq:diff_recur}.

The simplicity of the radial operator $\dop$ leads to some useful
phenomena. For example, making the variable substitution $r = \log s$
we can expand powers $\dop^n$ directly without recourse to the product
rule, which is desirable when applying automatic differentiation for
the purposes of computing \eqref{eq:diff_recur}. Also, while we often
find that evaluating potential basis functions $\Phi_{nl}$ is tricky
at small $r$ due to floating-point error, through Ramanujan's master
theorem we can directly read off the Taylor coefficients around $r=0$
from the Mellin transform of $\Phi_{nl}$, the $k$-th coefficient being
proportional to $p_{nl}\!\left(\imagi(k+1/2)\right)$
\citep{Lilley2025}. \refedit{Another trick made possible by the
  homogeneity of $\dop$ with respect to $r$ is the identity
  $p_{nl}(\dop)f(a \vec{r}) = p_{nl}(\imagi(a\partial_a +
  5/2))f(a\vec{r})$, which means application of the operator can be
  moved out from under an integral sign.}

\subsubsection{\refedit{Example: exponential basis set}}\label{sec:spherical_examples}

\refedit{%
  A fairly simple illustrative example, not found in the literature
  above. The density functions resemble Slater-type (hydrogenic)
  atomic orbitals. The zeroth-order radial density and potential
  functions are
\begin{equation}
  \rho_{0l} = r^l \: \expe^{-r}, \qquad
  \Phi_{0l} = \frac{-4\pi}{2l+1} \left[ r^{-l-1} \upgamma(3+2l,r) + r^l (1+r) \expe^{-r} \right],
\end{equation}
with $\upgamma(a,z)$ a lower incomplete Gamma function. The
index-raising polynomial is a continuous symmetric Hahn polynomial
$p_{nl}(s) = p_n(s/2; 1/4 + l/2, 7/4 + l/2)$ \dlmf{18.19}, giving the
normalisation constant
\begin{equation}
N_{nl} = \frac{4^{l+2} \Gamma \left(n+l+1/2\right) \Gamma (n+l+2)^2 \Gamma \left(n+l+7/2\right)}{n! (2 n+2 l+3) \Gamma (n+2 l+3)}.
\end{equation}
Each higher-order density basis function depends on two generalised
Laguerre polynomials
\begin{align}\label{eq:slater_rhonl_result}
  \rho_{nl} &=  A_{nl} r^l \expe^{-r} \left[ (2n+2l+1) L_n^{(2l+4)}(2r)-(2n+2l+5) L_{n-2}^{(2l+4)}(2r) \right], \\ \nonumber
  A_{nl} &= \frac{\imagi^n (l+2)_n (l+1/2)_n}{(2l+1) (2l+3)_n},
\end{align}
and each higher-order potential basis function (normalised as
$\hat{\Phi}_{nl} = \Phi_{nl}/B_{nl}$) is calculated via a recurrence
that depends on one Laguerre polynomial,
\begin{align}\label{eq:slater_phinl_result}
  \hat{\Phi}_{n+2,l} = \hat{\Phi}_{nl} + \frac{8\pi r^{l+2} \expe^{-r}}{2l+1} C_{nl} L_n^{(2l+4)}(2r), & \qquad %
  \hat{\Phi}_{1l} = \Phi_{0l} + \frac{8\pi r^l \expe^{-r}}{2l+1}(1+r), \\ \nonumber
  B_{nl} = \frac{(-\imagi)^n (l+1/2)_n (l+2)_n}{n!}, & \qquad %
  C_{nl} = \frac{(-1)^n n! (2n+2l+5)}{(l+2)(2l+3)(2l+5)_n}.
\end{align}
These expressions can be directly verified; for a derivation based on
generating functions, as well as the generalisation to the case of an
arbitrary $B$-function at zeroth-order, see \citet{Lilley2025}. This
example also displays the general pattern of basis sets whose
index-raising polynomial is classical: the density functions
themselves depend on $k$ classical polynomials (where $k$ does not
depend on $n$), whereas the potential functions satisfy a three-term
inhomogeneous recurrence relation whose non-homogeneous term
involves $p$ classical polynomials with $1 \leq p < k$. %
}

\begin{figure}
  \centering
  \includegraphics[width=0.5\textwidth]{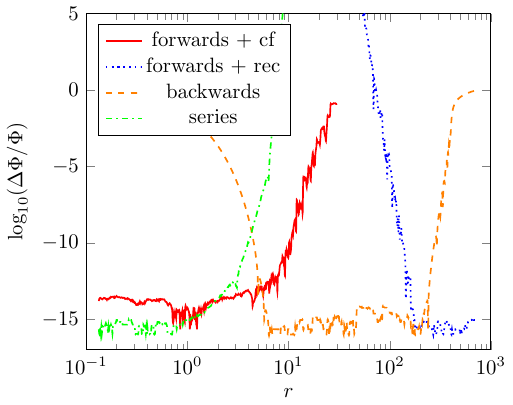}
  \caption{\refedit {The different regimes of accuracy when computing
      the \textit{Slater-adapted} $\Phi_{nl}$
      \eqref{eq:slater_phinl_result} (here with $n=30$ and
      $l=40$). The `forwards $+$ cf' uses forwards recursion for
      $\Phi_{nl}$ and a continued fraction for $\Phi_{0l}$ (`forwards
      $+$ rec' is similar but with a recurrence relation for
      $\Phi_{0l}$); `backwards' uses backwards recursion for
      $\Phi_{nl}$ with $100$ terms, and `series' the Taylor series
      with $90$ terms.  On the $y$-axis, $\Delta \Phi$ is the absolute
      difference between the given method and the `exact'
      (arbitrary-precision arithmetic) value, so
      $|\log_{10}(\Delta \Phi/\Phi)|$ is the number of correct decimal
      digits.}  }\label{fig:slater_accuracy}

\end{figure}

\refedit{%
  For such potential functions a naive evaluation is not necessarily
  stable. There are two ways to interpret
  \eqref{eq:slater_phinl_result}: either a forwards recurrence
  relation for each $\Phi_{nl}$ in terms of $\Phi_{n-2,l}$, or a
  backwards\footnote{\refedit{This is analogous to the classical
      `Miller's method' in the evaluation of Bessel functions.}}
  relation for $\Phi_{nl}$ in terms of $\Phi_{n+2,l}$ (with the
  assumption that $\Phi_{Nl} \approx 0$ for sufficiently large
  $N$). Therefore when $n$ is high there are three regimes to
  consider: small $r$, for which a Taylor expansion is appropriate;
  intermediate $r$, for which the forwards recurrence is stable; and
  large $r$, for which the backwards recurrence is
  stable. Furthermore, increasing $l$ tends to shrink the middle
  region, and at very large $r$ the $1/r$ behaviour of the
  zeroth-order term dominates over the higher-order terms, so the
  forwards recursion is again the most accurate. Therefore the correct
  evaluation strategy depends on stitching together these regions,
  which we have illustrated in Fig.~\ref{fig:slater_accuracy}. Use of
  a backwards recurrence for $\Phi_{nl}$ or $\grad \Phi_{nl}$ at high
  $n$ (for part of the range of $r$) is a general requirement for
  spherical polar basis sets whose index-raising polynomials are
  classical \citep{LSEE,Lilley2025}.
}

\refedit{%
  Note that these concerns do not apply to the density basis functions
  $\rho_{nl}$, as they depend on a fixed number of classical
  orthogonal polynomials, which have stable recurrences \citep{AS};
  nor is stability a concern in those very special cases wherein both
  potential and density basis functions depend on only a single
  classical polynomial (for example those given in
  Sec.~\ref{sec:cylindrical_examples}, \ref{sec:spheroidal_plummer}
  and App.~\ref{sec:more_prolate_spheroidal}, as well as the simpler
  spherical polar examples of \citet{CB73,HO92,Zh96}).%
}%

\subsection[Cylindrical polar]{Cylindrical polar \coordref{II}{2}}\label{sec:cyl}
Exceptionally in this section we take $R = \sqrt{x^2 + y^2}$.
\begin{flalign*}
\text{\textit{Coords:}} \quad & q_1 = R, \quad q_2 = \varphi, \quad q_3 = z, \quad x = R\cos{\varphi}, \quad y = R\sin{\varphi} && \\
                              & h_1 = 1, \quad h_2 = -R, \quad h_3 = -1, \quad f_1 = R, \quad f_2 = -1, \quad g = R^2 && \\
\text{\textit{Operators:}} \quad & L_1 = R^{-1}\partial_R + \partial_R^2, \quad L_2 = R^{-2}\partial_\varphi^2, \\
                              & T_1^* = \{\kop_3^*, \pop_3\} - \dop^{*2} + 1/4 = -\left(R^2 \partial_R^2 + R \partial_R + R^2 \partial_z^2\right), \quad T_2 = \jop_3, \quad T_3 = \pop_3
\end{flalign*}
This coordinate system is simultaneously rotational and cylindrical;
to make our formalism work we have to treat $z$ as the `azimuthal'
coordinate $q_3$ (and flip the sign of $h_3$ compared to Morse \&
Feshbach). The eigenfunctions involve imaginary-order modified Bessel
functions $K_{\imagi \alpha}(kR)$ (real-valued for real arguments),
\begin{align}
\phi_{\alpha m k}(R,\varphi,z) &= \frac{-4\pi}{\alpha^2 + m^2} K_{\imagi \alpha}\!\left(kR\right) \: \expe^{\imagi kz + \imagi m \varphi}, \\ \nonumber
\Psi_{\alpha m k}(R,\varphi,z) &= R^{-2} \: K_{\imagi \alpha}\!\left(kR\right) \: \expe^{\imagi kz + \imagi m \varphi},
\end{align}
which satisfy
\begin{equation}
  \left(\substack{T_1\\T_2\\T_3}\right) \Psi_{\alpha m k} = \left(\substack{\alpha^2\\m\\k}\right) \Psi_{\alpha m k},
  \qquad\qquad \left(\substack{T_1^*\\T_2^*\\T_3^*}\right) \phi_{\alpha m k} = \left(\substack{\alpha^2\\m\\k}\right) \phi_{\alpha m k}.
\end{equation}
The orthogonality relation for the eigen-densities is
\begin{equation}
\left\langle \Psi_{\alpha m k}, \Psi_{\beta\mu\kappa} \right\rangle = \frac{16 \pi^5}{\alpha \sinh{(\pi\alpha)}(\alpha^2 + m^2)} \delta_{m\mu} \delta(k - \kappa) \delta(\alpha - \beta),
\end{equation}
and the expansion coefficient of $\varrho$ in the $\Psi_{\alpha km}$
basis (the transform \eqref{eq:Ulm_transform}) is
\begin{equation}
  \left\langle \varrho, \Psi_{\alpha m k} \right\rangle = \frac{2\pi^3}{\alpha \sinh{(\pi\alpha)}(\alpha^2 + m^2)} \: \int_0^\infty \frac{dR}{R} \int_{-\infty}^\infty \! dz \int_0^{2\pi} \! d\varphi \: K_{\imagi \alpha}\!\left(kR\right) \: \expe^{-\imagi k z -\imagi m \varphi} \: \varrho(R,\varphi,z),
\end{equation}
which is a combination of a Kontorovich-Lebedev transform in the
radial variable \dlmf{10.43(v)}, a Fourier transform in $z$, and a
Fourier series in $\varphi$. In order to construct a conventional
basis set we must restrict the range of $z$ so that $k$ takes on
discrete values. Another possibility is to abandon completeness in the
$z$-direction and integrate the $k$-dependence out of the weight
function $\omega_{mk}$. We provide an example of such a cylindrical
basis set with `fixed' $z$-dependence in
Sec.~\ref{sec:cylindrical_examples}.

\refedit{%
Verifying the Coulomb resolution \eqref{eq:completeness_relation} is
similar to the spherical polar case: we take the completeness relation
for the polynomials $p_{nmk}$ \eqref{eq:poly_complete}, and then
calculate
\begin{align}\label{eq:cyl_complete}
  \!\!\!\!\!\!\!\!\!\!\!\!\!\!\!\! \sum_{km} \sum_{n=0}^\infty \frac{\Phi_{nmk}(\vec{r}) \overline{\Phi_{nmk}(\vec{r}^\prime)}}{N_{nm}}
  &= \frac{2}{\pi^3} \int_{0}^\infty \!\!\! dk \: \expe^{\imagi k(z - z^\prime)} \!\! \sum_{m=-\infty}^\infty \!\! \expe^{\imagi m (\varphi - \varphi^\prime)} \int_0^\infty \!\!\! d\alpha \frac{\alpha \sinh{(\pi\alpha)}}{\alpha^2 + m^2} K_{\imagi \alpha}(k R) K_{\imagi \alpha}(k R^\prime) \\ \nonumber
  &= \frac{1}{\pi} \int_{0}^\infty \!\! dk \: \expe^{\imagi k(z - z^\prime)} \!\! \sum_{m=-\infty}^\infty \!\! \expe^{\imagi m (\varphi - \varphi^\prime)} \times \begin{cases} I_m(k R^\prime) K_m(k R), & R^\prime < R \\ I_m(kR) K_m(kR^\prime), & R < R^\prime \end{cases},
\end{align}
which again is the appropriate Green's function expansion in
cylindrical coordinates \citep[Eq.~(3.148)]{Jackson2ed} (justfying the
sum-to-integral replacement in $k$ by the Poisson summation formula,
and then evaluating the integral over $\alpha$ using \bmp{12.1(8)}).%
}

\subsubsection{Example: basis set with fixed $z$-dependence}\label{sec:cylindrical_examples}

\refedit{%
If one is willing to impose finite boundary conditions in the $z$
direction, then the basis elements will be orthogonal and complete in
the $z$ direction, forming a Fourier series in $z$ (each element
depends on a single wavenumber $k$ and is proportional to
$\expe^{\pm \imagi k z}$).%
}

However, given that in practice the vertical profile of galactic discs
is fairly uniform \refedit{\citep{Kruit82}}, a more intriguing
possibility is to fix the $z$-dependence of the zeroth-order potential
and density but retain infinite boundary conditions in $z$. The
zeroth-order now does not have to be separated\footnote{Some separated
  potential and density factors have been tabulated in
  \citet{Ea96}. However the important difference is that Earn's method
  assumes the potential form
  $ \Phi_m = J_m(kR) \: \expe^{\imagi m \varphi} f(z)$, so that the
  corresponding density factor is $f'' - k^2 f$; whereas separated
  potentials in our case would take the form
  $ \Phi_m = K_{\imagi \alpha}(kR) \: \expe^{\imagi m \varphi} f(z)$,
  so the corresponding density factor is
  $f'' - (r^{-2}(\alpha^2 + m^2) - k^2)f$.} in $R$ and $z$, it merely
has to have a tractable Fourier transform in $z$.

Given the $k$-dependent weight function $\omega_{km}(\alpha^2)$
\eqref{eq:subspace_integration}, the new weight function with the
$z$-component `integrated out' is simply
\begin{equation}\label{eq:omega_cylindrical}
  \omega_m(\alpha^2) = \int_{-\infty}^{\infty} \! dk \:\: \omega_{km}(\alpha^2).
\end{equation}
Or more explicitly, if the zeroth-order density is
$\varrho_{0m} = f_m(R,z) \: \expe^{\imagi m \varphi}$, then the new
weight function is
\begin{align}\label{eq:omega_cylindrical_explicit}
  \omega_m(\alpha^2) &= \frac{\alpha \sinh{(\pi\alpha)}}{\alpha^2 + m^2} \int_{-\infty}^{\infty} dk \: \left| {\hat f}_m(\alpha,k) \right|^2, \\ \nonumber
  {\hat f}_m(\alpha,k) &= \int_0^{\infty} R \: dR \: K_{\imagi \alpha}(kR) \: \int_{-\infty}^{\infty} dz \: \expe^{-\imagi k z} \: f_m(R,z).
\end{align}
A basis set constructed in this manner will be orthogonal over all
space, but \emph{not} complete in the $z$-direction. It would probably
therefore be most useful for systems that have rapid and uniform
fall-off with increasing $|z|$. \refedit{Furthermore if the density is
  separable in $R$ and $z$, as is realistic for galactic discs
  \citep{Cuddeford93}, then the integral for ${\hat f}_m(\alpha,k)$ in
  \eqref{eq:omega_cylindrical_explicit} can be split into two
  one-dimensional integrals.}

\begin{figure}
  \centering
  \includegraphics[width=0.9\textwidth]{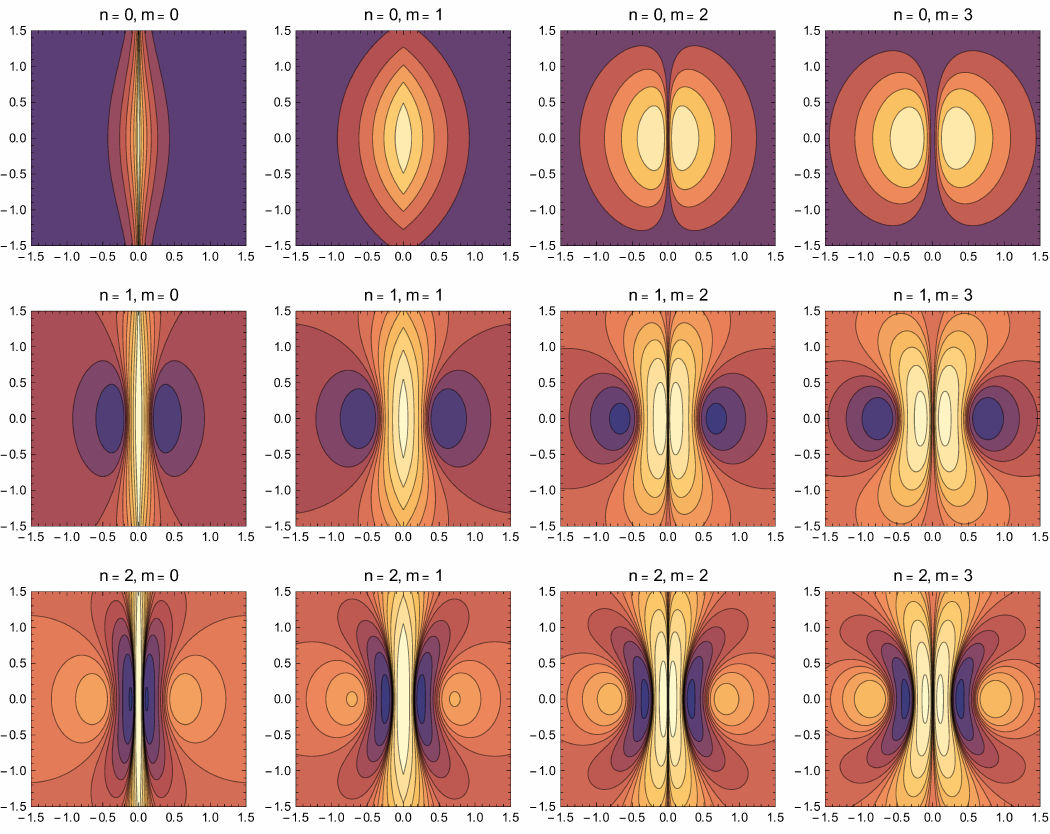}
  \includegraphics[width=0.9\textwidth]{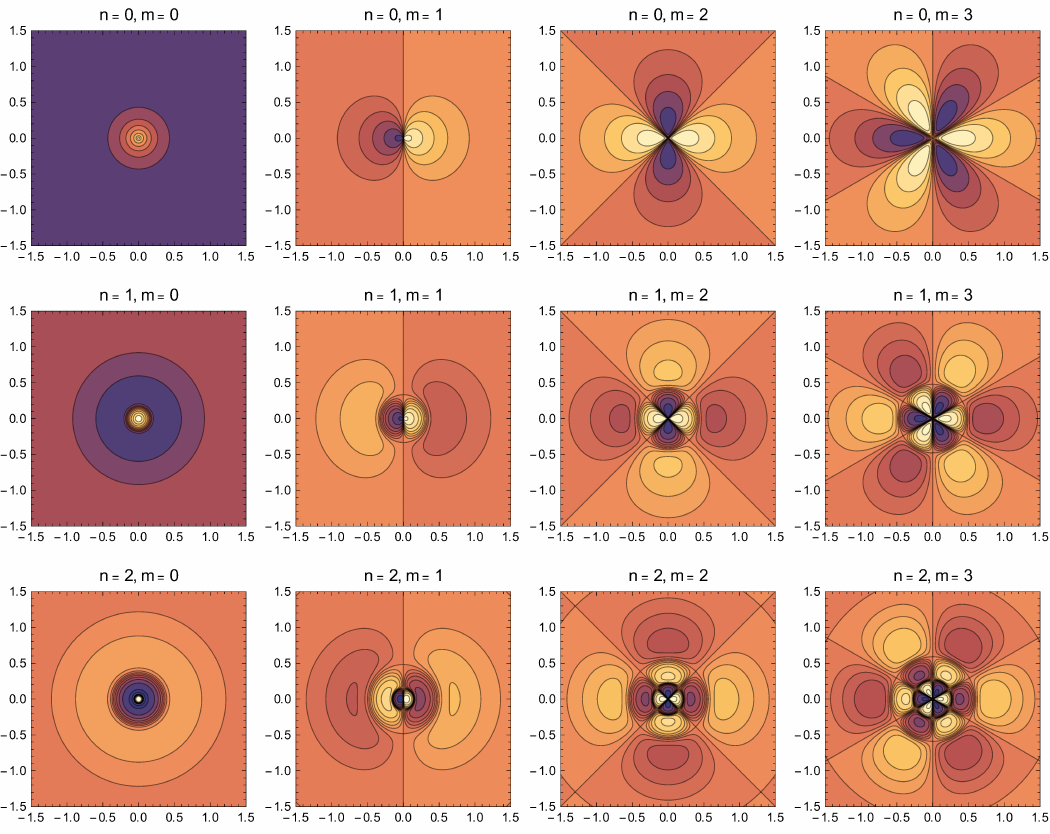}
  \caption{\refedit{Cross-sections of some `prolate Hernquist' density basis
    functions \eqref{eq:cyl_hq}, viewed along the $y$-axis (top $3$
    rows) and $z$-axis (bottom $3$ rows). The indices shown are
    $n = 0 \ldots 2$ and $m = 0 \ldots 3$, with scale parameter
    $b = 1$.}}\label{fig:cylindrical_hernquist}
\end{figure}

As a simple \refedit{(albeit somewhat non-physical)} example, we can
take the following `prolate Hernquist' model\footnote{A special case
  of the `prolate Plummer' model, \refedit{the dynamics of which were
    studied} in \citet{Jiang2007}.} (elongated along the $z$-axis by a
parameter $b$) with zeroth-order potential and density
\begin{align}\label{eq:cyl_hq_zeroth}
  \Phi_{0m}(R,\varphi,z) &= \frac{-4\pi}{b \: (2|m|+1)^2} \frac{R^{|m|} \: \expe^{\imagi m \varphi}}{\left( (R+b)^2 + z^2 \right)^{{|m|}+1/2}}, \\ \nonumber
  \varrho_{0m}(R,\varphi,z) &= \frac{R^{{|m|}-1} \: \expe^{\imagi m \varphi}}{\left( (R+b)^2 + z^2 \right)^{{|m|}+3/2}}.
\end{align}
The Fourier transform in $z$ \prud{Vol.~1, 2.3.5(7)}, the
Kontorovich-Lebedev transform in $R$ \gradsh{6.583}, and the final
$k$-integral \eqref{eq:omega_cylindrical} \gradsh{6.576(4)} can all be
found in closed-form, so we obtain the weight function
\begin{equation}\label{eq:omega_cyl_hernquist}
  \omega_m(\alpha^2) = \frac{\pi}{2^{4{|m|}+6} \Gamma({|m|}+3/2)^2 b^{2{|m|}+3}} \left| \frac{ \Gamma(1/2+ \imagi\alpha)^2 \: \Gamma({|m|}+\imagi\alpha) \: \Gamma({|m|}+1+\imagi\alpha) }{\Gamma(2\imagi\alpha)} \right|^2.
\end{equation}
This is proportional to the weight for the Wilson polynomial
$W_n(\alpha; 1/2,1/2,|m|,|m|+1)$ \dlmf{18.25}, so we take that as our
index-raising polynomial $p_{nm}(\alpha^2)$. Then, computing
$p_{nm}(T_1^*) \Phi_{0m}$, we find that each basis function is proportional to
a single Jacobi polynomial $P_n^{(2 |m|,0)}(\chi)$,
\begin{align}\label{eq:cyl_hq}
  \chi &= \frac{R^2-6bR+b^2+z^2}{(R+b)^2+z^2}, \\ \nonumber
  \Phi_{nm}(R,\varphi,z) &= n! \: (|m|+1/2)_n^2 \: P_n^{(2 |m|,0)}(\chi) \: \Phi_{0m}(R,\varphi,z), \\ \nonumber
  \varrho_{nm}(R,\varphi,z) &= n! \: (|m|+3/2)_n^2 \: P_n^{(2 |m|,0)}(\chi) \: \varrho_{0m}(R,\varphi,z). %
\end{align}
where the expression for $\Phi_{nm}$ follows from a sum over terms of
the form
\begin{equation}\label{eq:cyl_hq_term}
  \left|\left(|m| + \imagi\sqrt{T_1^*}\right)_k\right|^2 \Phi_{0m} = \prod_{j=0}^{k-1} \left( (|m|+j)^2 + T_1^*\right) \Phi_{0m} = \frac{2^{2k} (|m|+1/2)_k^2 \: R^k \: \Phi_{0m}}{\left((R+b)^2 + z^2\right)^{k}}.
\end{equation}
Some representative density basis functions are shown in
Fig.~\ref{fig:cylindrical_hernquist}.

We can apply \eqref{eq:cyl_hq_term} to find new zeroth-orders, and
hence find the weight functions for some similar models, such as the
more physically-plausible Plummer-like density
\begin{equation}
(T_1 + (m+1)^2) \varrho_{0m} \propto R^m \expe^{\imagi m \varphi} \left( (R+b)^2 + z^2) \right)^{-m-5/2},
\end{equation}
although it appears that our original choice \eqref{eq:cyl_hq_zeroth}
is the only one of this form that gives a classical index-raising
polynomial. More generally we note that the presence of a double
integral
\eqref{eq:omega_cylindrical}--\eqref{eq:omega_cylindrical_explicit} in
the definition of the weight function means that this style of basis
set is somewhat more reliant on convenient integrals being found in
reference tables.

The Kontorovich-Lebedev-Fourier transform relation between Wilson and
Jacobi polynomials implied by
\eqref{eq:omega_cylindrical_explicit}--\eqref{eq:cyl_hq} does not
appear to correspond to any of the results of \citet{Koornwinder1985},
and therefore may be new.

\subsection[Prolate spheroidal]{Prolate spheroidal \coordref{VIII}{6}}\label{sec:pro_sph}
\begin{flalign*}
\text{\textit{Coords:}} \quad & q_1 = \eta, \quad q_2 = \vartheta, \quad q_3 = \varphi && \\
                              & x = b \sinh\eta \sin\vartheta \cos\varphi, \quad y = b \sinh\eta \sin\vartheta \sin\varphi, \quad z = b \cosh\eta \cos\vartheta && \\
                              & h_1 = -h_2 = b\sqrt{\cosh^2\eta - \cos^2\vartheta}, \quad h_3 = -b \sinh\eta \sin\vartheta, \\
                              & f_1 = b \sinh\eta, \quad f_2 = -\sin\vartheta, \quad g = b^2\left(\sinh^2\eta + \sin^2\vartheta\right) && \\
                              \text{\textit{Operators:}} \quad & L_1 = g^{-1}\left(\coth\eta\partial_\eta + \partial_\eta^2\right), \quad L_2 = g^{-1}\left(\cot\vartheta\partial_\vartheta + \partial_\vartheta^2\right), \\
& T_1^* = \dop^{*2} - b^2 \pop_3^2 - b^2\sin^2\!\vartheta \Lap + 1/4 = -\left(\partial_\eta^2 + \coth\eta \partial_\eta + \cosech^2\eta \partial_\varphi^2\right), \\
& T_2^* = -\dop^{*2} + b^2\pop_3^2 - b^2 \sinh^2\!\eta \Lap - 1/4 = -\left(\partial_\vartheta^2 + \cot\vartheta \partial_\vartheta + \cosec^2\vartheta \partial_\varphi^2\right), \quad T_3 = \jop_3
\end{flalign*}
The eigenfunctions are\footnote{\refedit{The Legendre function
    $P_{\imagi \alpha - 1/2}^{(m)}\!(x)$ for integer $m$ and real
    $(\alpha, x)$ has been variously termed a Mehler, conical, or
    hyperboloid function.}}
\begin{align}\label{eq:prolate_spheroidal_eigenfunctions}
\phi_{\alpha l m}(\eta,\vartheta,\varphi) &= \frac{-4\pi}{\alpha^2 + (l+1/2)^2} \: P_{\imagi \alpha - 1/2}^{(m)}\!\left(\cosh\eta\right)\: P_l^{(m)}\!\left(\cos\vartheta\right) \: \expe^{\imagi m \varphi}, \\ \nonumber
\Psi_{\alpha l m}(\eta,\vartheta,\varphi) &= \frac{b^{-2}}{\sinh^{2}\!\eta + \sin^{2}\!\vartheta} \: P_{\imagi \alpha - 1/2}^{(m)}\!\left(\cosh\eta\right)\: P_l^{(m)}\!\left(\cos\vartheta\right) \: \expe^{\imagi m \varphi},
\end{align}
which satisfy
\begin{equation}
  \left(\substack{T_1\\T_2\\T_3}\right) \Psi_{\alpha l m} = \left(\substack{\alpha^2\\l(l+1)\\m}\right) \Psi_{\alpha l m},
  \qquad\qquad \left(\substack{T_1^*\\T_2^*\\T_3^*}\right) \phi_{\alpha l m} = \left(\substack{\alpha^2\\l(l+1)\\m}\right) \phi_{\alpha l m}.
\end{equation}
The constant $b$ is half the focal distance of the underlying ellipse;
when $b \to 0$ in $T_1$ we recover the spherical polar
case\footnote{To see that the same is true of $T_2^*$, first use
  \eqref{eq:sph_pol_lap_id} to rewrite it as
  $ T_2^* = \jop^2 + \left(r^2 - b^2\sinh^{2}\!\eta\right)\Lap + b^2
  \pop_3^2$, then note that $b^2\sinh^{2}\!\eta \to r^2$ as
  $b \to 0$.}, and when $b \to \infty$ the cylindrical
polar\footnote{In fact we first have to take $(x,y) \to (bx,by)$ and
  afterwards $b \to \infty$; then prolate spheroidal $b^{-2} T_j$
  become linear combinations of cylindrical polar $T_j$. The reduction
  to cylindrical polar coordinates is therefore somewhat non-trivial
  \mf{Ch.~5}.}.  The orthogonality relation is\footnote{Here it is
  important to include the spherical harmonic normalisation factor
  $J_{lm}$ explicitly.}
\begin{equation}
\left\langle \Psi_{\alpha lm}, \Psi_{\beta\lambda\mu} \right\rangle = 4 b^2 J_{lm} \frac{\left| \Gamma(\imagi \alpha + m + 1/2)\right|^2 \: \cosh{\!(\pi\alpha)}^2}{\alpha \sinh{(\pi\alpha)} \: \left(\alpha^2 + (l+1/2)^2\right)} \delta_{l \lambda} \delta_{m\mu} \delta(\alpha - \beta),
\end{equation}
so the expansion $\langle \Psi_{\alpha lm}, f \rangle$ (the transform
\eqref{eq:Ulm_transform}) combines a generalised Mehler-Fock transform
in $\eta$ with a spherical harmonic expansion in the angular
variables\footnote{Although note that the prolate spheroidal
  $\vartheta$ is \emph{not} the same as the spherical polar
  $\vartheta$, so the combination
  $P_l^{(m)}(\cos\vartheta) \: \expe^{\imagi m \varphi}$ is not equal
  to a standard spherical harmonic, although it serves the same role
  in the expansion formulas.}.  The inverse coordinate transformations
are
\begin{align}
\cosh^{2}\!\eta &= \frac{1}{2}\left(\frac{r^2}{b^2} + 1\right) + \frac{1}{2}\sqrt{1 + \frac{2(x^2+y^2-z^2)}{b^2} + \frac{r^4}{b^4}}, \\ \nonumber
\cos^{2}\!\vartheta &= \frac{1}{2}\left(\frac{r^2}{b^2} + 1\right) - \frac{1}{2}\sqrt{1 + \frac{2(x^2+y^2-z^2)}{b^2} + \frac{r^4}{b^4}}.
\end{align}
\refedit{%
The Coulomb resolution can be verified following \eqref{eq:completeness_relation},
  \begin{align}\label{eq:pro_complete}  \nonumber
\!\!\!\!\!\!\!\!\!\!\!\!\!\!\!\!\!\!\!\!\!\!\!\!\!\!\!\!\!\!\!\!\!\!\!\!\!\!\!\! \sum_{nlm} \frac{\Phi_{nlm}(\vec{r}) \: \overline{\Phi_{nlm}(\vec{r}^\prime)} }{ N_{nlm} }
    &= \frac{4 \pi^2}{b^2} \sum_{m} \expe^{\imagi m (\varphi - \varphi^\prime)} \!\int_0^\infty\!\! d\alpha \frac{\alpha \sinh{\!(\pi\alpha)} P^{(m)}_{\imagi \alpha - 1/2}(\cosh\eta) P^{(m)}_{\imagi\alpha - 1/2}(\cosh\eta^\prime)}{\cosh{\!(\pi\alpha)}^2 \: |\Gamma(\imagi \alpha + m + 1/2)|^2} \sum_{l} \frac{P_l^{(m)}(\cos\vartheta) P_l^{(m)}(\cos\vartheta^\prime)}{J_{lm} (\alpha^2 + (l+1/2)^2)} \\
    & \!\!\!\!\!\!\!\!\!\!\!\!\!\!\!\!\!\!\!\!\!\!\!\!\!\!\!\!\!\!\!\!\!\!\!\!\!\!\!\!\!\!\!\!\!\!\!\!\!\!\!\!\!\!\!\!\!\!\!\!\!\!\!\! = \frac{2\pi^2}{b^2} \sum_m \expe^{\imagi m (\varphi - \varphi^\prime)} \!\int_0^\infty\!\! d\alpha \frac{\alpha \sinh{\!(\pi\alpha)} \Gamma(\imagi\alpha + 1/2 - m)^2}{\cosh{\!(\pi\alpha)}^2 \Gamma(\imagi\alpha + 1/2 + m)^2} P_{\imagi\alpha - 1/2}^{(m)}(\cosh\eta) P_{\imagi\alpha - 1/2}^{(m)}(\cosh\eta^\prime) P_{\imagi\alpha - 1/2}^{(m)}(\cos\vartheta) P_{\imagi\alpha - 1/2}^{(m)}(\!-\!\cos\vartheta),
  \end{align}
  where we summed over $l$ \prud{Vol.~3, 6.5.5(2)} (taking care to
  include the $J_{lm}$ factor explicitly), and applied some reflection
  formulas for the gamma and Legendre functions \dlmf{5.5, \S 14.9}.
  The result matches an expression for the Coulomb potential first
  given by \citet[Eq.~27]{VanNostrand1954}.}

\subsubsection{Example: scale-free or Plummer basis set}\label{sec:spheroidal_plummer}

To construct a simple example of a prolate spheroidal basis set we can
take the zeroth-order potential and density
\begin{align}\label{eq:spheroidal_plummer_zeroth_order}
  \Phi_{0lm}(\eta,\vartheta,\varphi) &= \frac{-\sinh^{|m|}\!\eta}{\cosh^{l+|m|+1}\!\eta} \: P_l^{(m)}(\cos\vartheta) \: \expe^{\imagi m \varphi}, \\ \nonumber
  \varrho_{0lm}(\eta,\vartheta,\varphi) &= \frac{\pi}{b^2}(l+|m|+1)(l+|m|+2)\frac{\sinh^{|m|}\!\eta}{\cosh^{l+|m|+3}\!\eta} \: \frac{P_l^{(m)}(\cos\vartheta) \: \expe^{\imagi m \varphi}}{\sinh^2\!\eta + \sin^2\!\vartheta}.
\end{align}
This looks like a \citet{Plu1911} model in the plane $z=0$, but
becomes scale-free as $b \to 0$ because there is only one length scale
in the model, which is proportional to the focal distance $b$. Given
that $T_1$ and $T_2$ are invariant under a simultaneous rescaling
$(b,x,y,z) \mapsto (kb,kx,ky,kz)$, this means that the same invariance
applies to all the potential-density pairs, but this \refedit{may limit
the usefulness of the density basis functions} in applications\footnote{Because changing
  the focal distance ($=$ scale-length) also re-scales the radial
  oscillations, the effective axis ratio of the model cannot be
  adjusted. Of course, choosing a zeroth-order with a more elaborate
  $b$-dependence avoids this issue.}.

\begin{figure}
  \centering
  \includegraphics[width=0.9\textwidth]{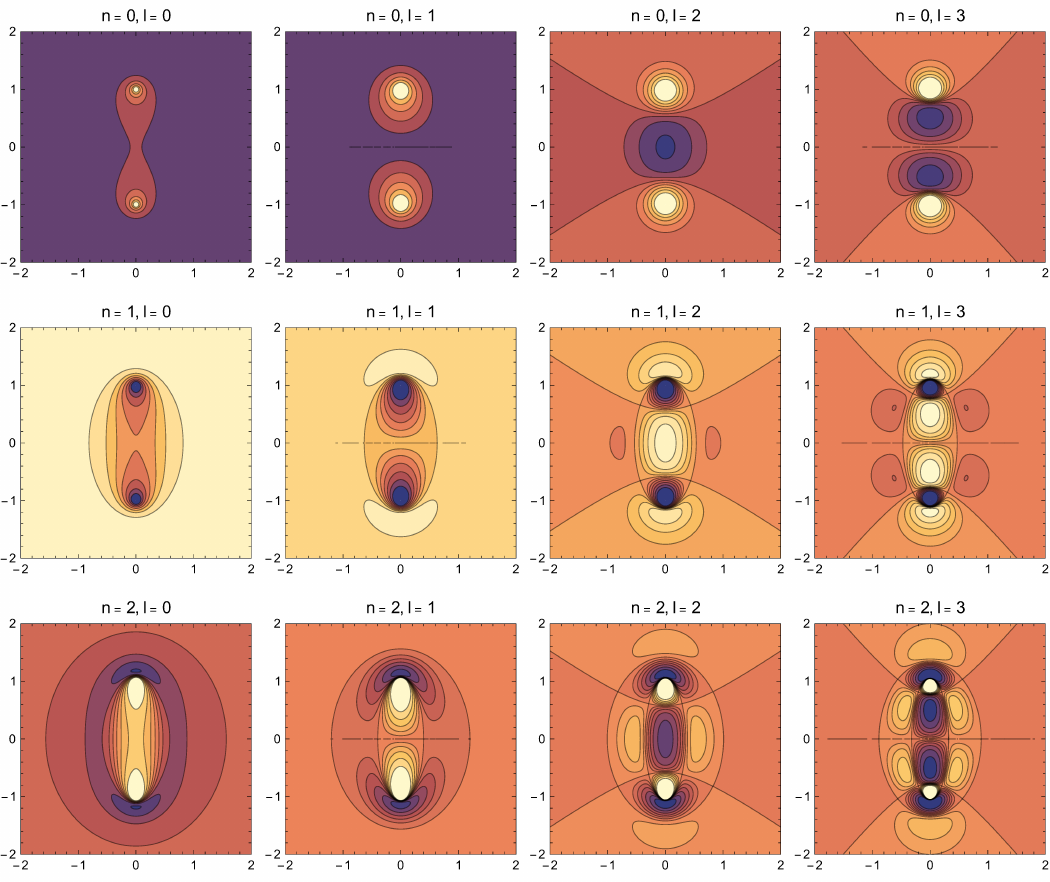}
  \includegraphics[width=0.9\textwidth]{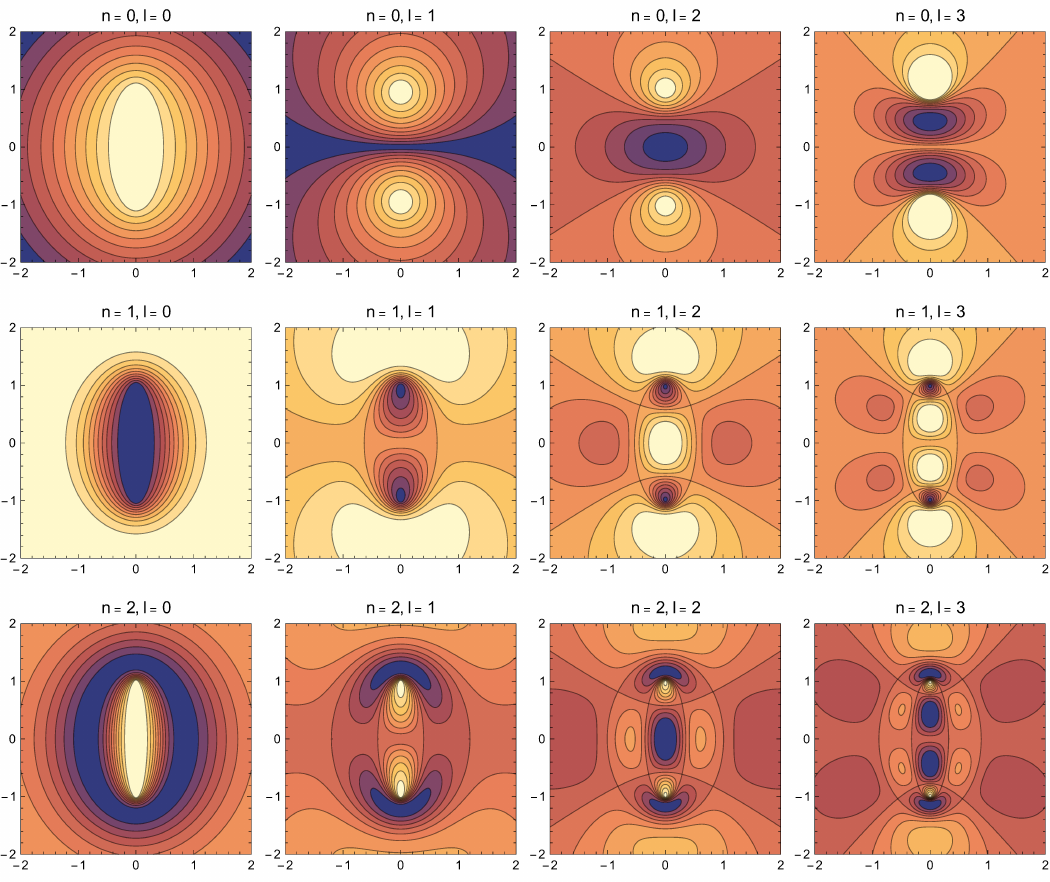}
  \caption{\refedit{Cross-sections of some prolate spheroidal
      `Plummer/scale-free' \eqref{eq:spheroidal_plummer_basis} density
      (top $3$ rows) and potential (bottom $3$ rows) basis functions,
      viewed along $y$-axis. The indices shown are $n=0\ldots 2$,
      $l=0\ldots 3$, $m=0$ and the focal distance is
      $b=1$.}}\label{fig:spheroidal_plummer}
\end{figure}

We can evaluate the inverse Mehler-Fock transform of $\Phi_{0lm}$
\gradsh{7.132(7)}, so the polynomial weight function becomes
\begin{align}\label{eq:spheroidal_plummer_weight}
  \omega_{lm}(\alpha^2) &= \frac{\alpha \sinh(\pi\alpha)}{4 \cosh{\!(\pi\alpha)}^2} \left| \frac{\imagi \alpha + l + 1/2}{\Gamma(\imagi \alpha + |m| + 1/2)} \right|^2 \: \left| \int_0^\infty d\eta \: \sinh\eta \: \Phi_{0lm}(\eta) \: P_{\imagi \alpha - 1/2}^{(m)}(\cosh \eta) \right|^2 \\ \nonumber
                      &= \frac{2^{2l+2|m|-2}}{\pi^3 (|m|+l)!^2} \left| \frac{\Gamma\!\left(\frac{\imagi \alpha + l + 5/2}{2}\right) \: \Gamma\!\left(\frac{\imagi \alpha + l + 1/2}{2}\right) \: \Gamma\!\left(\frac{\imagi \alpha + |m| + 3/2}{2}\right) \: \Gamma\!\left(\frac{\imagi \alpha + |m| + 1/2}{2}\right) }{\Gamma(\imagi \alpha)} \right|^2,
\end{align}
which is proportional to the weight function for the Wilson polynomial
$W_n(x;a_l,b_l,c_m,d_m)$ with parameters
$(a_l,b_l,c_m,d_m) = (l/2 + 5/4, l/2 + 1/4, |m|/2 + 3/4, |m|/2 + 1/4)$
\dlmf{18.25}, so for the index-raising polynomial we have
\begin{align}
  \!\!\!\!\!\!\!\!\!\!\!\! p_{nlm}(\alpha^2) &= W_n\!\left( \alpha^2/4; a_l,b_l,c_m,d_m\right) \\ \nonumber
  \!\!\!\!\!\!\!\!\!\!\!\!                                   &= \left(l + \frac{3}{2}\right)_n \left(\frac{l+|m|}{2} + 2\right)_n \left(\frac{l+|m|}{2} + \frac{3}{2}\right)_n
  \hypergeom{4}{3}{-n,n+l+|m|+\frac{5}{2},\frac{l}{2} + \frac{5}{4} + \frac{\imagi \alpha}{2},\frac{l}{2} + \frac{5}{4} - \frac{\imagi \alpha}{2}}{  l + \frac{3}{2}, \frac{l+|m| + 4}{2}, \frac{l+|m| + 3}{2}}{1}.
\end{align}
The normalisation constant is
\begin{equation}
  N_{nlm} = -\intddd \: \Phi_{nlm} \: \overline{\varrho_{nlm}} = \frac{2^{2l+2|m|-3}}{\pi^4 (l+|m|)!^2} h_n(a_l,b_l,c_m,d_m)\: J_{lm},
\end{equation}
where $h_n(a_l,b_l,c_m,d_m)$ is the normalisation constant for the
Wilson polynomials \dlmf{18.25.5}. Working out the form of
$\Phi_{nlm} = p_{nlm}(T_1^*)\Phi_{0lm}$ explicitly\footnote{Aided by
  the identity
  $(a + \imagi x)_n (a - \imagi x)_n = \prod_{j=0}^{n-1} \left(
    (a+j)^2 + x^2\right)$, as well as some expressions
  \eqref{eq:prolate_operator_identity_0}--\eqref{eq:prolate_operator_identity_1}
  found in App.~\ref{sec:more_prolate_spheroidal}.}, we find that each
basis function is proportional to a single Jacobi polynomial,
\begin{align}\label{eq:spheroidal_plummer_basis}
  \xi = \frac{\sinh^2\!\eta - 1}{\sinh^2\!\eta + 1}, \quad & \quad A_{nlm} = n! \left((l+|m| + 2)/2\right)_{\! n} \! \left((l+|m|+1)/2\right)_{\! n}, \\ \nonumber
  \Phi_{nlm}(\eta,\vartheta,\varphi) &= A_{nlm} \: P_n^{(l+1/2,|m|)}\!\left(\xi\right) \: \Phi_{0lm}(\eta,\vartheta,\varphi), \\ \nonumber
  \varrho_{nlm}(\eta,\vartheta,\varphi) &= A_{n,l+2,m} \: P_n^{(l+1/2,|m|)}\!\left(\xi\right) \: \varrho_{0lm}(\eta,\vartheta,\varphi).
\end{align}
Some of these are plotted in Fig.~\ref{fig:spheroidal_plummer}. The
basis set resembles a generalisation of \citet{CB73}'s original
Plummer-adapted basis set to prolate spheroidal coordinates.

Any basis set constructed in this way suffers from a defect (observed
by \citet{Ro96}) whereby the unavoidable
$\left(\sinh^2\!\eta + \sin^2\!\vartheta\right)^{-1}$ factor in the
density causes singularities at the points of the `focal needle'
$\refedit{\sinh\eta} = \sin\vartheta = 0$\footnote{These singularities
  are absorbed by the volume element
  $d^{3\,}\!\vec{r} = b^3 \left(\sinh^2\!\eta +
    \sin^2\!\vartheta\right) \sinh\eta \sin\vartheta d\eta d\vartheta
  d\varphi$.}.  We provide another related example of a prolate
spheroidal basis set (which resembles an isochrone model) in
App.~\ref{sec:more_prolate_spheroidal}, as well as some discussion
about further generalisation.

\subsection[Oblate spheroidal]{Oblate spheroidal \coordref{IX}{7}}\label{sec:obl_sph}
\begin{flalign*}
\text{\textit{Coords:}} \quad & q_1 = \eta, \quad q_2 = \vartheta, \quad q_3 = \varphi && \\
                              & x = b \cosh\eta \sin\vartheta \cos\varphi, \quad y = b \cosh\eta \sin\vartheta \sin\varphi, \quad z = b \sinh\eta \cos\vartheta && \\
                              & h_1 = -h_2 = b\sqrt{\sinh^2\eta + \cos^2\vartheta}, \quad h_3 = -b \cosh\eta \sin\vartheta, \\
                              & f_1 = b \cosh\eta, \quad f_2 = -\sin\vartheta, \quad g = b^2\left(\sinh^2\eta + \cos^2\vartheta\right) && \\
                              \text{\textit{Operators:}} \quad & L_1 = g^{-1}\left(\coth\eta\partial_\eta + \partial_\eta^2\right), \quad L_2 = g^{-1}\left(\cot\vartheta\partial_\vartheta + \partial_\vartheta^2\right), \\
                              & T_1^* = \dop^{*2} + b^2 \pop_3^2 + b^2 \sin^{2}\!\vartheta \Lap + 1/4 = -\left(\partial_\eta^2 + \tanh\eta \partial_\eta - \sech^2\eta \partial_\varphi^2\right), \\
                              & T_2^* = -\dop^{*2} - b^2\pop_3^2 - b^2\cosh^{2}\!\eta \Lap - 1/4 = -\left(\partial_\vartheta^2 + \cot\vartheta \partial_\vartheta + \cosec^2\vartheta \partial_\varphi^2\right), \quad T_3 = \jop_3
\end{flalign*}
The eigenfunctions could use either of the solutions to Legendre's
equation,
\begin{align}
\phi_{\alpha l m}(\eta,\vartheta,\varphi) &= \frac{-4\pi}{\alpha^2 + (l+1/2)^2} \: \left(\mathlarger{\mathlarger{\substack{P\\Q}}}\right)_{\imagi \alpha - 1/2}^{(m)}\!\left(\imagi\sinh\eta\right) \: P_l^{(m)}\!\left(\cos\vartheta\right) \: \expe^{\imagi m \varphi}, \\ \nonumber
\Psi_{\alpha l m}(\eta,\vartheta,\varphi) &= \frac{b^{-2}}{\sinh^{2}\!\eta + \cos^{2}\!\vartheta} \: \left(\mathlarger{\mathlarger{\substack{P\\Q}}}\right)_{\imagi \alpha - 1/2}^{(m)}\!\left(\imagi\sinh\eta\right) \: P_l^{(m)}\!\left(\cos\vartheta\right) \: \expe^{\imagi m \varphi}.
\end{align}
If the first-kind functions
$P^{(m)}_{\imagi \alpha - 1/2}(\imagi \sinh\eta)$ are chosen a
Mehler-Fock transform results, but a straight-forward analogy to the
prolate case is complicated by the imaginary argument. If instead the
second-kind functions
$Q^{(m)}_{\imagi \alpha - 1/2}(\imagi \sinh\eta)$ are chosen, we can
apply a Whipple formula \dlmf{14.9(iv)} to rewrite them in terms of
first-kind functions with real argument,
\begin{equation}
Q_{\imagi \alpha - 1/2}^{(m)}\!\left(\imagi\sinh\eta\right) = -\imagi \sqrt{\frac{\pi}{2}} \frac{\expe^{\imagi m \pi} \: \Gamma(\imagi \alpha + m + 1/2)}{\sqrt{\cosh \eta}} P_{m-1/2}^{(-\imagi \alpha)}\!\left(\tanh\eta\right).
\end{equation}
The integral transform associated with these functions\footnote{The
  canonical name for $P_{m-1/2}^{(\imagi \alpha)}(x)$ would be a
  \emph{toroidal function} of imaginary order.} seems little-studied,
although several orthogonality relations with respect to $\alpha$ can
be found in \citet{Bielski2013} -- but the simplest orthogonality
relation only applies if we assume $\alpha > 0$ and (fictitiously)
take $\eta \in (-\infty,\infty)$. We have so far been unable to find a
simple example of an oblate spheroidal basis set (analogous to the
prolate results) using this method, but numerical construction is
still an option.

On the plane $z=0$ the situation simplifies, as $\eta = 0$ and we are
naturally restricted to the disc $R<b$
\citep[App.~F]{qian_thesis}. Hence only eigenfunctions of $T_2$ are
required, which already have a discrete eigenvalue $l$ and therefore
no tri-diagonalisation is necessary. This case was first studied by
\citet{Hunter1963}, who used eigen-densities proportional to
$(\cos{\vartheta})^{-1} P_l^{(m)}(\cos{\vartheta}) \expe^{\imagi m
  \varphi}$, a result which was then generalised to give a family of
finite-disc basis sets with varying power-law behaviour by
\citet{Ka76}, using a variation of his logarithmic spiral
technique. This suggests that a theory of basis sets of finite extent
might be encompassed by the present operator-based formalism, the
details of which are yet to be worked out.

\subsection[Parabolic]{Parabolic \coordref{VII}{8}}\label{sec:par_pol}
\begin{flalign*}
\text{\textit{Coords:}} \quad & q_1 = \eta, \quad q_2 = \sigma, \quad q_3 = \varphi && \\
                              & x = \eta \sigma \cos\varphi, \quad y = \eta \sigma \sin\varphi, \quad z = (\eta^2 - \sigma^2)/2 && \\
                              & h_1 = h_2 = \sqrt{\eta^2 + \sigma^2}, \quad h_3 = -\eta\sigma, \quad f_1 = \eta, \quad f_2 = \sigma, \quad g = \eta^2 + \sigma^2 && \\
\text{\textit{Operators:}} \quad & L_1 = g^{-1}\left(\frac{1}{\eta}\partial_\eta + \partial_\eta^2\right), \quad L_2 = g^{-1}\left(\frac{1}{\sigma}\partial_\sigma + \partial_\sigma^2\right), \\
                              & T_1^* = \frac{1}{2}\left(\{\pop_1,\jop_2\} - \{\pop_2,\jop_1\} + \{\pop_3,\dop^*\} - g\Lap\right) = -\left(\eta^2 \partial^2_\eta + \eta \partial_\eta + \partial^2_\varphi \right), \\
                              & T_2^* = -\frac{1}{2}\left(\{\pop_1,\jop_2\} - \{\pop_2,\jop_1\} + \{\pop_3,\dop^*\} + g\Lap\right) = -\left(\sigma^2 \partial^2_\sigma + \sigma \partial_\sigma + \partial^2_\varphi \right), \quad T_3 = \jop_3
\end{flalign*}
\refedit{The eigenfunctions use Bessel functions $J_m(x)$,}
\begin{align}
\phi_{k q m}(\eta,\sigma,\varphi) &= \frac{-4\pi}{k^2 + q^2} \: J_{m}\!\left(k \eta\right) \: J_{m}\!\left(q \sigma\right) \: \expe^{\imagi m \varphi}, \\ \nonumber
\Psi_{k q m}(\eta,\sigma,\varphi) &= \frac{1}{\eta^2 + \sigma^2} \: J_{m}\!\left(k \eta\right) \: J_{m}\!\left(q \sigma\right) \: \expe^{\imagi m \varphi},
\end{align}
which satisfy
\begin{equation}
  \left(\substack{T_1\\T_2\\T_3}\right) \Psi_{k q m} = \left(\substack{k^2\\q^2\\m}\right) \Psi_{k q m},
  \qquad\qquad \left(\substack{T_1^*\\T_2^*\\T_3^*}\right) \phi_{k q m} = \left(\substack{k^2\\q^2\\m}\right) \phi_{k q m},
\end{equation}
and the orthogonality relation is
\begin{equation}
\left\langle \Psi_{k q m}, \Psi_{\kappa \zeta \mu} \right\rangle = \frac{4\pi \: \delta_{m\mu} \: \delta(k - \kappa) \: \delta(q - \zeta)}{kq(k^2 + q^2)}.
\end{equation}
However, to form a viable basis set one of the coordinates
$(\eta,\sigma)$ must be restricted to a finite range; the transform
$f \mapsto \left\langle \Psi_{k q m}, f \right\rangle$ then involves a
Hankel transform in one coordinate and a Fourier-Bessel series in the
other.

\subsection[Bispherical]{Bispherical \coordref{p.~665}{not listed separately}}\label{sec:bis}
\begin{flalign*}
\text{\textit{Coords:}} \quad & q_1 = \eta, \quad q_2 = \vartheta, \quad q_3 = \varphi && \\
                              & x = \frac{b\sin\vartheta \cos\varphi}{\cosh\eta - \cos\vartheta}, \quad y = \frac{b\sin\vartheta \sin\varphi}{\cosh\eta - \cos\vartheta}, \quad z = \frac{b\sinh\eta}{\cosh\eta - \cos\vartheta} && \\
                              & h_1 = -h_2 = \frac{b}{\cosh\eta - \cos\vartheta}, \quad h_3 = \frac{-b \sin\vartheta}{\cosh\eta - \cos\vartheta}, \quad f_1 = 1, \quad f_2 = -\sin\vartheta \\
                              & R = \frac{\sqrt{b}}{\sqrt{\cosh\eta - \cos\vartheta}}, \quad g = R^4 && \\
                              \text{\textit{Operators:}} \quad & L_1 = g^{-1}\left(\frac{-\sinh\eta}{\cosh\eta - \cos\vartheta}\partial_\eta + \partial_\eta^2\right), \quad
                              L_2 = g^{-1}\left(\frac{1-\cosh\eta\cos\vartheta}{\sin\vartheta\left(\cosh\eta - \cos\vartheta\right)}\partial_\vartheta + \partial_\vartheta^2\right), \\
                              & T_1^* = \frac{1}{b}\kop_3^* - \frac{b}{2}\pop_3, \quad
                              T_2^* = -\left(\frac{1}{b}\kop_3^* - \frac{b}{2}\pop_3\right)^2 - g\Lap - \frac{1}{4}, \quad T_3 = \jop_3
\end{flalign*}
The eigenfunctions are
\begin{align}
\phi_{s l m}(\eta,\vartheta,\varphi) &= \frac{4\pi}{\sqrt{b}} \: \frac{\sqrt{\cosh\eta - \cos\vartheta}}{s^2 + (l+1/2)^2} \: P_{l}^{(m)}\!\left(\cos\vartheta\right)\: \expe^{\imagi s \eta + \imagi m \varphi}, \\ \nonumber
\Psi_{s l m}(\eta,\vartheta,\varphi) &= b^{-5/2} \left(\cosh\eta - \cos\vartheta \right)^{5/2} \: P_{l}^{(m)}\!\left(\cos\vartheta\right)\: \expe^{\imagi s \eta + \imagi m \varphi}.
\end{align}
This gives rise to a Fourier transform in the radial variable $\eta$
and a spherical harmonic expansion in the angular
variables\footnote{Once again, in these coordinates $\vartheta$ is
  \emph{not} the same as the spherical polars' $\vartheta$.}. Upon
substituting $u = \expe^\eta$ the radial transform becomes a Mellin
transform in $u$, resembling the spherical polar case. In fact,
bispherical coordinates are a conformal transformation of spherical
polar coordinates, which means that any spherical polar basis set can
be transformed into a bispherical basis set according to
\begin{align}\label{eq:spherical_bispherical_trans}
\varrho_{nl}(r) Y_{lm}(\vartheta, \varphi) \mapsto \left(\cosh{\eta} - \cos{\vartheta}\right)^{5/2} \: \left[r^{5/2}\varrho_{nl}(r)\right]_{r = a \expe^\eta}\: Y_{lm}(\vartheta,\varphi), \\ \nonumber
\Phi_{nl}(r) Y_{lm}(\vartheta, \varphi) \mapsto \left(\cosh{\eta} - \cos{\vartheta}\right)^{1/2} \: \left[r^{1/2}\Phi_{nl}(r)\right]_{r = a \expe^\eta}\: Y_{lm}(\vartheta,\varphi)
\end{align}
(up to some factors of $a$ and $b$). Note the coordinate $\vartheta$
changes meaning between the left and right sides. To write the new basis
functions in Cartesian coordinates use
\begin{align}
  \cosh{\eta} &= \frac{x^2 + y^2 + z^2 + b^2}{\sqrt{(x^2 + y^2 + (z+b)^2)(x^2 + y^2 + (z-b)^2)}}, \quad \expe^\eta = \sqrt{\frac{x^2 + y^2 + (z+b)^2}{x^2 + y^2 + (z-b)^2}}, \\ \nonumber
  \cos{\vartheta} &= \frac{x^2 + y^2 + z^2 - b^2}{\sqrt{(x^2 + y^2 + (z+b)^2)(x^2 + y^2 + (z-b)^2)}}.
\end{align}
The transformation \eqref{eq:spherical_bispherical_trans} generically
maps a single-centred spherical polar basis set into a
\emph{two}-centred \emph{bi}spherical basis set\footnote{Compare the
  `A $\leftrightarrow$ B' (Kelvin) transformation of \citet{LSE},
  which affects only the radial variable (and maps the set of
  double-power law models into itself), and hence the transformed
  basis sets remain single-centred. }, whose nodes along the $z$-axis
are somewhat concentrated around the poles $z = \pm b$.
The parameter $b$ is half of the separation between the poles, and the
parameter $a$ adjusts the relative mass concentration between them
(equal at $a = 1$, biased towards the $z = -b$ pole as $a \to 0$,
biased towards the $z = +b$ pole as $a \to \infty$).

As an example, if we take the two-parameter $(\nu,\alpha)$ family of
basis sets of \citet{LSE} and apply this bispherical transformation,
we obtain a new family of basis sets whose density functions have
\refedit{one cusp proportional to
  $|\vec{r} - b \hat{\vec z}|^{\nu/\alpha-2}$ as
  $\vec{r} \to (0,0,b)$, a second cusp proportional to
  $|\vec{r} + b \hat{\vec z}|^{1/\alpha-2}$ as $\vec{r} \to (0,0,-b)$,
  and a fixed power-law tail $r^{-5}$ as
  $r \to \infty$}\footnote{Geometrically this is because the conformal
  transformation \eqref{eq:spherical_bispherical_trans} moves the
  inner power law from $r=0$ to $z=-b$, and the outer one from
  $r=\infty$ to $z=b$.}. To fix the same power law dependency at each
cusp we must therefore set $\nu = 1$; and to have no cusps at all
(cores) we must have $\nu = 1$ and $\alpha = 1/2$, which produces a
Plummer model at zeroth order centred at $z = (1-a^2)/(1+a^2)$. Such
basis sets may be useful for modelling interacting dark matter haloes.

\begin{figure}
  \centering
  \includegraphics[width=0.9\textwidth]{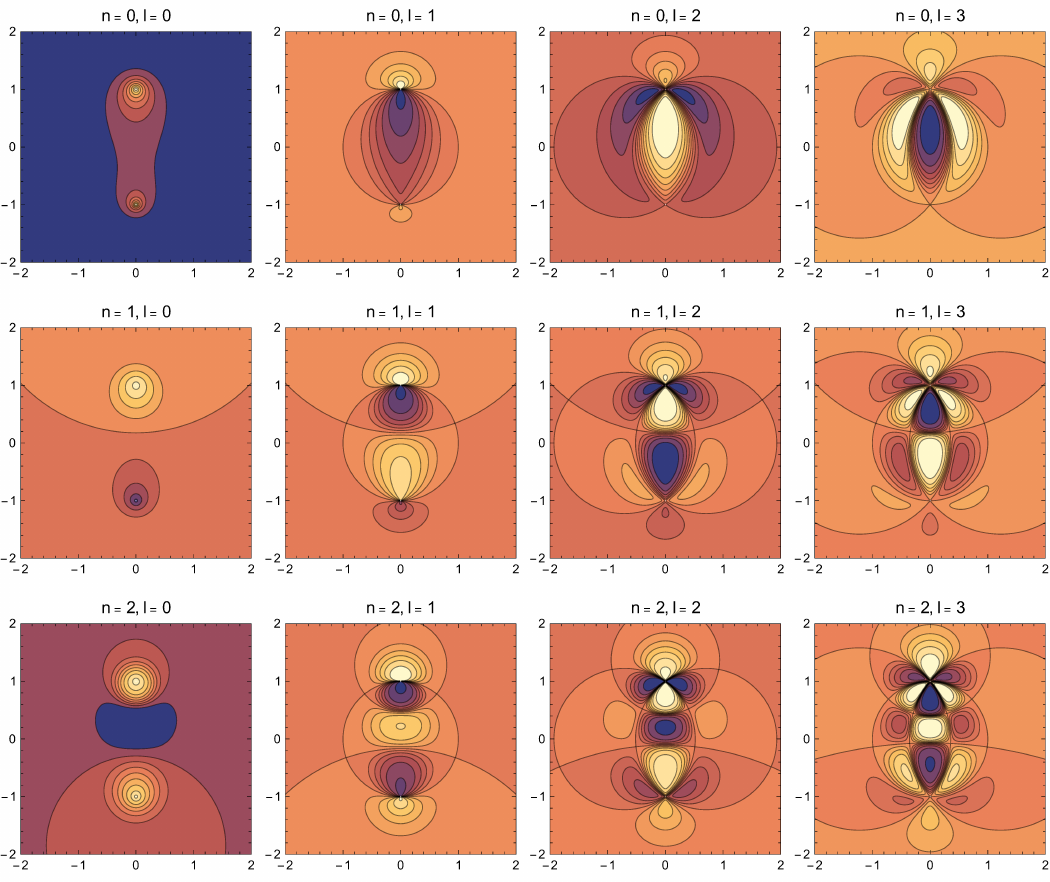}
  \includegraphics[width=0.9\textwidth]{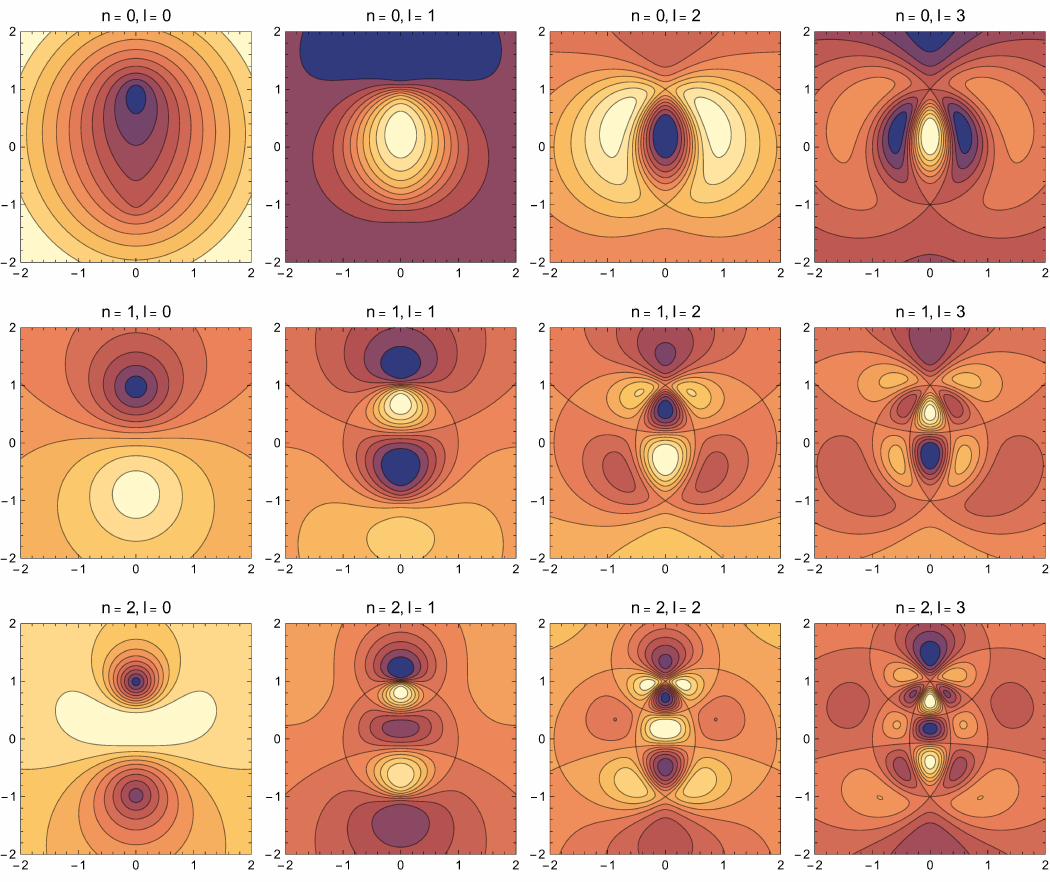}
  \caption{\refedit{Taking the well-known
    \citet{HO92} results and transforming them according to
    \eqref{eq:spherical_bispherical_trans} to produce `bisphericalised' Hernquist-Ostriker
    density (top $3$ rows) and potential (bottom $3$ rows) basis
    functions, all viewed along the $y$-axis. The indices shown are
    $n = 0 \ldots 2$, $l = 0 \ldots 3$ and $m=0$; the mass
    concentration parameter is $a = 0.7$ (unequal cusp masses),
    and the cusp separation is $b = 1$.}
  }\label{fig:bispherical_hernquist}
\end{figure}

\begin{figure}
  \centering
  \includegraphics[width=0.9\textwidth]{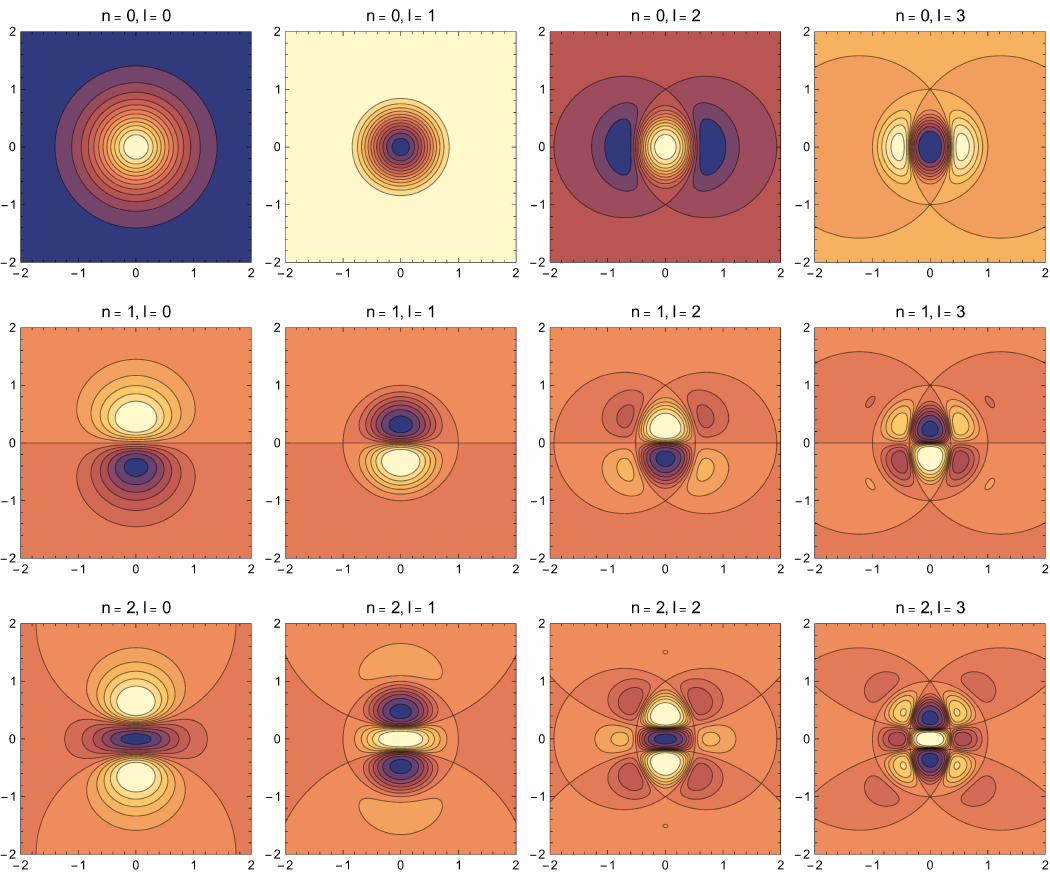}
  \includegraphics[width=0.9\textwidth]{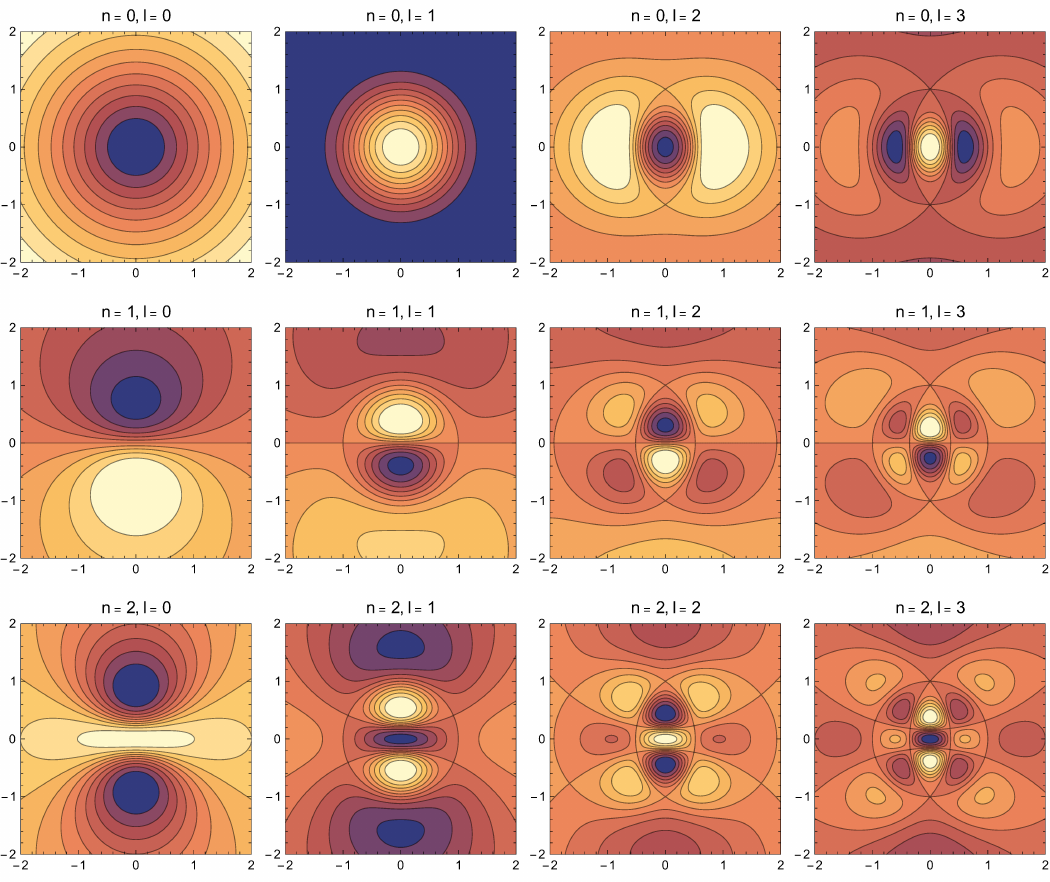}
  \caption{\refedit{Similar to Fig.~\ref{fig:bispherical_hernquist} but instead
    `bisphericalising' the \citet{CB73} basis functions and choosing
    the parameters $a = b = 1$, producing a bispherical basis with an
    exact Plummer model as zeroth-order, but with bispherical higher
    order terms. Showing density (top 3 rows) and potential (bottom 3
    rows) basis functions for $n = 0 \ldots 2$, $l = 0 \ldots 3$ and
    $m=0$, all viewed along $y$-axis.}
  }\label{fig:bispherical_plummer}
\end{figure}

See Fig.~\ref{fig:bispherical_hernquist} for some representative
contour plots of `bisphericalised' \citet{HO92} density basis
functions, where both poles have \refedit{a cusp proportional to
  $|\vec{r} \pm b \hat{\vec z}|^{-1}$, and with the cusps'} masses
set to be slightly unequal. See Fig.~\ref{fig:bispherical_plummer} for
the limiting case of a `bisphericalised' Plummer model, which is
exactly a standard Plummer model at zeroth order and whose higher
order terms represent bispherical (rather than spherical)
perturbations.

Of note is another interesting link between bispherical coordinates
and basis sets on the plane $z=0$ provided by \citet{Hu80}, who
re-derived \citet{CB72}'s disc basis set using the method of
images. It is likely that other disc basis sets have an alternative
derivation in this manner.

\subsection[Toroidal]{Toroidal \coordref{p.~666}{17}}\label{sec:tor}
\begin{figure}
  \centering
  \includegraphics[width=\textwidth]{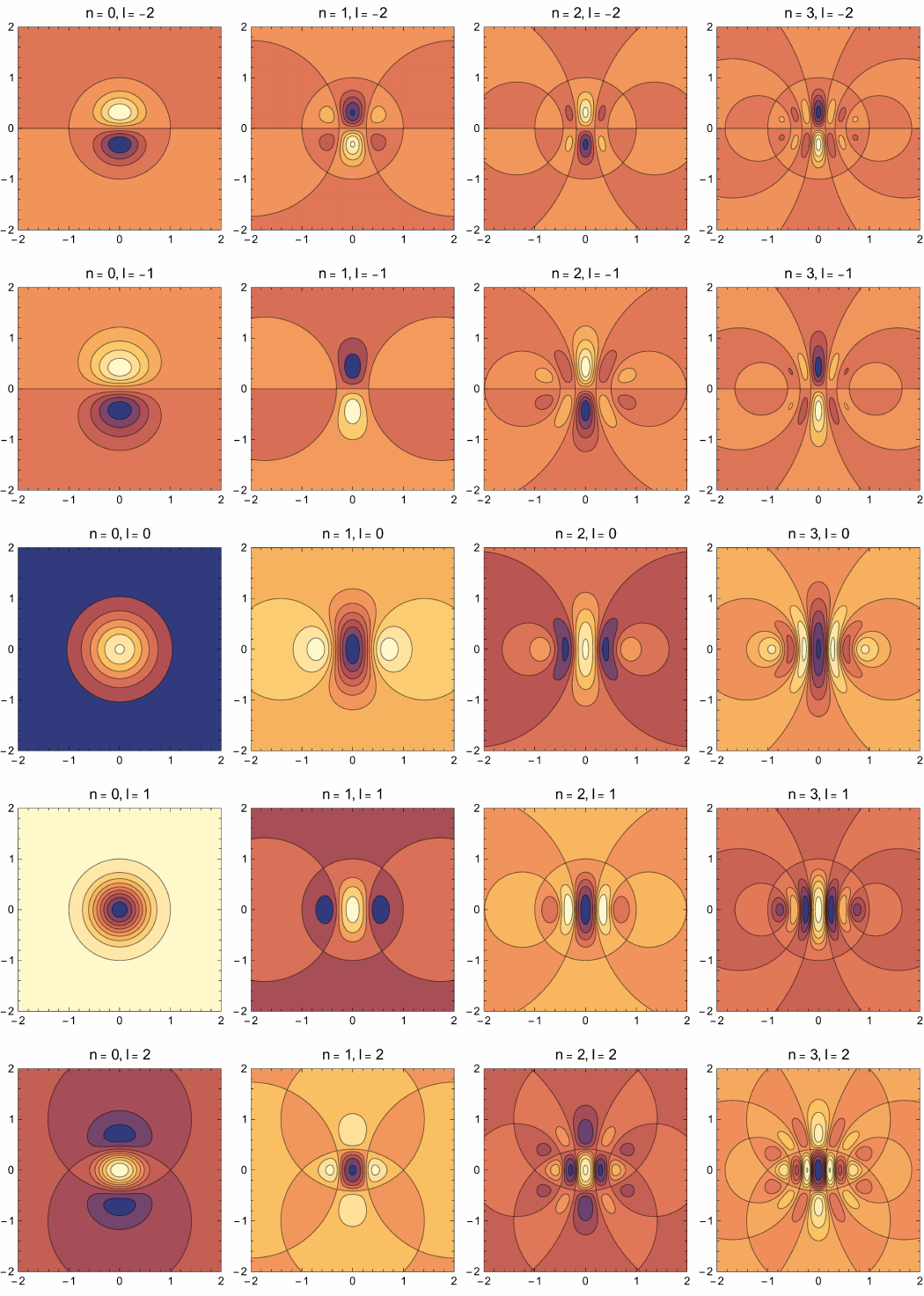}
  \caption{\refedit{Applying the `prolate spheroidal $\leftrightarrow$
    toroidal' transformation of Sec.~\ref{sec:tor} to the
    `Plummer/scale-free' basis functions of
    Sec.~\ref{sec:spheroidal_plummer} gives yet another basis set with
    an exact Plummer model at zeroth order, but with the higher-order
    terms now representing toroidal perturbations. Here we show
    density basis functions for $n=0\ldots 2$ (across the page),
    $l = -2 \ldots 2$ (down the page), $m=0$ and $b=1$, all viewed
    along the $y$-axis.}}\label{fig:toroidal_plummer}
\end{figure}

\begin{flalign*}
\text{\textit{Coords:}} \quad & q_1 = \eta, \quad q_2 = \vartheta, \quad q_3 = \varphi && \\
                              & x = \frac{b\sinh\eta \cos\varphi}{\cosh\eta - \cos\vartheta}, \quad y = \frac{b\sinh\eta \sin\varphi}{\cosh\eta - \cos\vartheta}, \quad z = \frac{b\sin\vartheta}{\cosh\eta - \cos\vartheta} && \\
                              & h_1 = -h_2 = \frac{b}{\cosh\eta - \cos\vartheta}, \quad h_3 = \frac{-b \sinh\eta}{\cosh\eta - \cos\vartheta}, \quad f_1 = \sinh\eta, \quad f_2 = -1, \\
                              & R = \frac{\sqrt{b}}{\sqrt{\cosh\eta - \cos\vartheta}}, \quad g = R^4 && \\
                              \text{\textit{Operators:}} \quad & L_1 = g^{-1}\left(\frac{1-\cosh\eta\cos\vartheta}{\cosh\eta - \cos\vartheta}\partial_\eta + \partial_\eta^2\right), \quad L_2 = g^{-1}\left(\frac{-\sin\vartheta}{\cosh\eta - \cos\vartheta}\partial_\vartheta + \partial_\vartheta^2\right), \\
                              & T_1^* = -\left(\frac{1}{b}\kop_3^* + \frac{b}{2}\pop_3\right)^2 - g\Lap + \frac{1}{4}, \quad T_2^* = \frac{1}{b}\kop_3^* + \frac{b}{2}\pop_3, \quad T_3 = \jop_3
\end{flalign*}
The eigen-potentials and -densities are
\begin{align}
  \phi_{\alpha lm}(\eta,\vartheta,\varphi) &= \frac{-4\pi}{\sqrt{b}} \: \frac{\sqrt{\cosh\eta - \cos\vartheta}}{\alpha^2 + l^2} P_{\imagi \alpha - 1/2}^{(m)}\!\left(\cosh\eta\right) \expe^{\imagi l \vartheta + \imagi m \varphi}, \\ \nonumber
  \Psi_{\alpha lm}(\eta,\vartheta,\varphi) &= b^{-5/2} \left(\cosh\eta - \cos\vartheta\right)^{5/2} P_{\imagi \alpha - 1/2}^{(m)}\!\left(\cosh\eta\right) \expe^{\imagi l \vartheta + \imagi m \varphi},
\end{align}
which satisfy the eigenvalue equations
\begin{equation}
  \left(\substack{T_1\\T_2\\T_3}\right) \Psi_{\alpha lm} = \left(\substack{\alpha^2\\l\\m}\right) \Psi_{\alpha lm},
  \qquad\qquad \left(\substack{T_1^*\\T_2^*\\T_3^*}\right) \phi_{\alpha lm} = \left(\substack{\alpha^2\\l\\m}\right) \phi_{\alpha lm}.
\end{equation}
As in the prolate spheroidal case, the expansion in $\Psi_{slm}$ gives
rise to a generalised Mehler-Fock transform; the associated
orthogonality relation is
\begin{equation}
\left\langle \Psi_{\alpha lm}, \Psi_{\beta\lambda\mu} \right\rangle = \frac{4 J_{lm}}{b} \: \frac{\left| \Gamma(\imagi \alpha + m + 1/2)\right|^2 \: \cosh{\!(\pi \alpha)}^2}{\alpha \sinh{(\pi \alpha)} \: \left(\alpha^2 + l^2\right)} \delta_{l \lambda} \delta_{m\mu} \delta(\alpha - \beta).
\end{equation}
Similarly to the spherical $\!\leftrightarrow\!$ bispherical case, we
can map prolate spheroidal to toroidal basis sets simply by
relabelling the coordinates, changing $l \mapsto |l| - 1/2$, and
adjusting the pre-factors. Performing this transformation on the
prolate spheroidal `Plummer/scale-free' basis set of
Sec.~\ref{sec:spheroidal_plummer} produces a new `toroidal Plummer'
basis set whose zeroth-order is an ordinary spherical Plummer model
but where the higher-order elements represent toroidal perturbations
(see Fig.~\ref{fig:toroidal_plummer}). Performing the same
transformation on the prolate spheroidal `isochrone' basis set of
App.~\ref{sec:more_prolate_spheroidal} produces a basis set whose
zeroth-order density has a ring-shaped singularity on the $z=0$ plane
at $R = b$. Such toroidal basis sets may be appropriate for
investigating the stability to toroidal perturbations of spherical
equilibrium models, or the stability of toroidal objects themselves,
such as ring-shaped galaxies \citep{Wong1974}.

\section{Discussion}\label{sec:discussion}

In practice the most useful basis sets -- those with compact
expressions for the actual potential and density basis functions --
appear to correspond to cases in which the associated index-raising
polynomial $p_{n}(x)$ is of Meixner-Pollaczek, continuous Hahn or
Wilson type. The modern description of these polynomials involves
their weight functions and recurrence relations \dlmf{18.19}, similar
to any other classical orthogonal polynomial. But historically
speaking, their development begins with the work of
\citet{Bateman1934}, who defined his polynomials $F_n(x)$ (a
specialisation of the continuous Hahn polynomials) using the property
\begin{equation}
  F_n(d/dx) \sech(x) = \sech(x) P_n(\tanh(x)),
\end{equation}
where $P_n(x)$ is a Legendre polynomial. After a change of variable
this strongly resembles the various expressions of
$p_{nl}(\dop)\varrho_{0l}$ that we find for the known spherical polar
basis sets. A general overview of the history of the continuous Hahn
and Meixner-Pollaczek polynomials can be found in \citet{Koe1996}, in
which can also be found an integral transform relationship between the
continuous Hahn and Jacobi polynomials\footnote{Further discussion on
  Fourier, Mellin and Mehler-Fock transforms of classical orthogonal
  polynomials can be found in
  \citet{Koornwinder1985,Koornwinder1989,Ismail2012}.}  (as well as
Meixner-Pollaczek and Laguerre polynomials after taking a suitable
limit). The various statements about index-raising polynomials in
\citet{Lilley2023} turn out to be basically slight variations of this
transform relationship. Together with the results in the current work
that involve Wilson polynomials (which are one level `higher' than
continuous Hahn polynomials in the standard hierarchy), this suggests
that the correspondence between polynomials, basis sets and symmetry
operators is fruitful and deserves further study.

We mentioned in passing above that our analysis could be repeated for
every coordinate system in the `thin-disc' ($z=0$) case. This was
studied in \citet{qian_thesis}\footnote{Some of which is published as
  \citet{Qi92,Qi93}.}, where for a few coordinate systems (notably
including prolate spheroidal coordinates) orthogonality conditions on
the disc were given analogously to those of Sec.~\ref{sec:coords}, but
without explicitly constructing the basis functions out of
operators. We summarise the operator-based approach on the disc as
follows: each Laplacian symmetry operator (listed in
App.~\ref{sec:algebra}) can be transferred onto the disc by making the
transformations
\begin{align}\label{eq:operator_disc_mapping}
  \pop_1, \jop_1 \mapsto \pop_1 \qquad  \pop_2, \jop_2 \mapsto \pop_2 \qquad \pop_3 \mapsto 0 \\ \nonumber
  \kop_1 \mapsto \mathpzc{k}_1 \qquad  \kop_2 \mapsto \mathpzc{k}_2 \qquad  \kop_3 \mapsto 0 \qquad \dop \mapsto \mathpzc{d}
\end{align}
where $\mathpzc{k}_1$, $\mathpzc{k}_2$ and $\mathpzc{d}$ are defined
in App.~\ref{sec:disc_operators} (similar transformations hold on the
$x=0$ and $y=0$ discs by permuting the indices). Terms in $f\Lap$
become
\begin{equation}
  f(x,y,z) \Lap \mapsto -f(x,y,0) \left( \pop_1^2 + \pop_2^2 \right).
\end{equation}
The disc operators resemble the original operators, so much of the
formalism for each coordinate system remains the same on the disc. The
eigen-densities and eigen-potentials on the disc differ in general
from their three-dimensional counterparts (although sometimes it is as
simple as setting $z=0$ in the original eigen-density
$\Psi_\lambda$)\footnote{This is why we wrote the energy inner
  product \eqref{eq:self} in terms of densities rather than potentials
  -- so that it generalises sensibly when taking
  $\varrho(x,y,z) = \sigma(x,y) \delta(z)$.}, as does the factor
$F_\eivec$ \eqref{eq:psi_orthog} that appears in the corresponding
integral transform.

The prototypical example is spherical polar coordinates: transferred
onto the disc we once again find a Mellin transform, but the
normalisation factor $F_\eivec$ becomes more complicated,
\begin{equation}
  \left| l+1/2 + \imagi s \right|^2 \: J_{lm} \longrightarrow
  \left| \frac{\Gamma\!\left(\frac{m+3/2+\imagi s}{2}\right)}{\Gamma\!\left(\frac{m + 1/2 + \imagi s}{2}\right)} \right|^2.
\end{equation}
What is remarkable is that this factor can in many cases still be
absorbed into the polynomial weight function in a natural way, giving
rise to fairly simple closed-form disk basis sets, of which
\citet{CB72} is the simplest example\footnote{Although it was
  \citet{CB72} who first wrote down this simple basis set whose
  zeroth-order is a Kuzmin/Toomre disc, it was \citet{Ka76} who
  re-derived it in terms of essentially a Mellin transform (his
  `logarithmic spirals' are the eigenfunctions $\Psi_{sm}$ on the
  disc), and \citet{Ao78} who identified that it could be written
  compactly in terms of Gegenbauer polynomials. A generalisation is
  given by \citet{Qi93}, and the index-raising polynomials are noted
  in \citet{Lilley2023}.}.

Cylindrical polar coordinates transferred to the disc give the same
eigenfunctions and operators as spherical polar coordinates; some
other coordinate systems produce different operators but result in the
same separation structure. For example with prolate spheroidal
coordinates we obtain the radial operator
$T_1 = \mathpzc{d}^{2} + b^2(\pop_1^2 + \pop_2^2)$, which gives rise
to a Mehler-Fock transform in the variable
$\cosh\eta = \sqrt{R^2/b^2 + 1}$. So although a different integral
transform is obtained, the same variables ($R$ and $\varphi$) appear
separated, so little is gained compared to the spherical polar
case. It would be worthwhile to classify what happens when each of the
coordinate systems listed in Sec.~\ref{sec:coords} is put onto the
$x=0$, $y=0$ or $z=0$ discs. This would hopefully clarify the
relationship with the alternative approaches that use oblate
spheroidal coordinates \citep{Hunter1963} and bispherical coordinates
\citep{Hu80}, for discs of finite and infinite extent respectively.

Our definition of the operators $T_j$ in Sec.~\ref{sec:operators} was
only justified heuristically. In fact \eqref{eq:Tj_def} is not a
\emph{unique} choice of three mutually-commuting quasi-commuting
operators. Trivially we can add any symmetric polynomial in one of the
operators to any of the other operators and retain these
properties. However there are also non-trivial alternative sets of
operators. For example,
\begin{equation}\label{eq:spherical_alternate}
  T_1 = \dop, \quad T_2 = \dop^2 - \{\kop_3,\pop_3\} - 1/4, \quad T_3 = \jop_3
\end{equation}
is an alternative choice of operators in spherical polar coordinates
(in this case only $T_2$ differs from \eqref{eq:Tj_def}). They obey
\begin{equation}
T_2^* + T_3^* = -r^2 \sin^{2}\!\vartheta \: \Lap,
\end{equation}
and have eigen-densities and eigen-potentials
\begin{equation}
  \phi_{plm} \propto r^{p} \: P_p^{(\imagi l)}(\cos\vartheta) \: \expe^{\imagi m \varphi},
  \quad \Psi_{plm} \propto \frac{r^{p-2}}{\sin^{2}\!\vartheta} \: P_p^{(\imagi l)}(\cos\vartheta) \: \expe^{\imagi m \varphi}.
\end{equation}
A similar (non-\eqref{eq:Tj_def}) choice of $T_j$ can be made in
prolate spheroidal coordinates. These particular alternative operators
are probably not so useful for producing basis sets, but a proper
classification of all the possible choices of $T_j$ would be
useful\footnote{Similar classification efforts can be found in
  e.g.~\citet{Makarov1967,Marchesiello2022}.}. It also bears
mentioning that the procedure of Sec.~\ref{sec:operators} does not
quite work for the more exotic cyclidic rotational $R$-separable
coordinate systems \citep{Bocher1894}. For some of the simpler systems
(e.g.~cardioid) just the definition of $u$ \eqref{eq:u_def} requires
modification, but for others the definition of $v_1$ and $v_2$
\eqref{eq:v_def} also breaks down.

From the perspective of work in galactic dynamics, it would naturally
be useful to find basis sets whose zeroth-orders are Stäckel
potentials. While this is always possible in the spherical case, the
spheroidal case suffers interference from the singular factor that
appears in the density functions
\eqref{eq:prolate_spheroidal_eigenfunctions}. However it may be
possible to control this factor, and this needs further investigation.

On a related note, missing from our analysis is the ellipsoidal
coordinate system. If suitable operators $T_j$ can be found, then by
analogy with the spheroidal cases the eigenfunctions would likely
contain a Lamé function of order $\imagi s - 1/2$, named
\emph{elliptic conal harmonics} by \citet{Hobson1891}. The resulting
integral transform is likely to be relatively intractable, however, as
little is known about integrals involving Lamé functions \citep{Arscott1964}.

In general, the source of the difficulty in handling Helmholtz
wavefunctions in spheroidal and ellipsoidal coordinates is their
non-hypergeometric nature, being instead solutions to Heun-type
equations. Hence it is (at least superficially) surprising that our
examples of prolate spheroidal basis sets bypass this difficulty,
instead remaining expressible in terms of hypergeometric functions.
It is possible that some of this good fortune may survive in the
ellipsoidal case.

A complete classification of all commuting triples of quadratic
Laplacian symmetry operators would also be desirable, although this
would require the solution of a very high-dimensional (and only
moderately sparse) system of bilinear equations (see discussion in
App.~\ref{sec:Sj_Tj_relationship}). In the present work we took
advantage of some of the known cases of separation of variables in
order to bypass this complexity. However a more general approach may
uncover viable (albeit exotic) non-separable coordinate systems.

\funding{The author acknowledges partial support from the European
  Research Council (ERC) under the European Union's Horizon 2020
  research and innovation programme under grant agreement No.~724857
  (Consolidator Grant ArcheoDyn).}

\data{Some \textsc{Mathematica} code implementing the methods and
  producing the figures of this paper is included in the repository at
  \url{https://github.com/ejlilley/basis}.}

\printbibliography[heading=bibintoc,title={References}]

\appendix

\section{Laplacian symmetry operators}\label{sec:algebra}

The operators $X$ such that $X \Lap = \Lap Y$ for some
operator $Y$ are the generalised symmetries of the Laplacian. Any
constant is a generalised symmetry; then for first-order operators
there is a basis $X_j$ of ten operators whose commutation relations
are isomorphic to $\mathfrak{so}(4,1)$.
For second-order operators we have
$55 = \properbinom{10}{2} + \properbinom{10}{1}$ symmetric
combinations of the form $\{X_i, X_j\} = X_i X_j + X_j X_i$, and
$\Lap f$ where $f$ is any \refedit{real-valued} function\footnote{Here
  and subsequently $\Lap f$ indicates an operator that operates to the
  right on another function, i.e.~$(\Lap f) g = \Lap(f
  g)$. \refedit{Hence $(\Lap f)^* = f \Lap$.}}. However, the
decomposition of a generic second-order symmetry operator
\begin{equation}
\mathcal{O} = \sum_{i,j} c_{i,j} \{X_i, X_j\} + \Lap f
\end{equation}
is non-unique, as there exist $20$ relations of the form
\begin{equation}
\mathcal{O}_1 - \mathcal{O}_2 = \Lap f_{1,2}
\end{equation}
where $\mathcal{O}_1$ and $\mathcal{O}_2$ are second-order operators
and $f_{1,2}$ is a function, possibly zero. If we identify two
operators (of any order) that differ by $Z \Lap$ for some third
operator or function $Z$ then we have the algebraic structure
described by \citet{Eastwood2002}, in which (up to equivalence) each
symmetry operator is canonically mapped to a different confomal
Killing tensor.

Crucially it turns out that all symmetric generalised Laplacian
symmetries satisfy the \textit{quasi-commutation} relation \refedit{\eqref{eq:quasi}},
\refedit{which implies that they are symmetric with respect to the
  energy inner product \eqref{eq:self}}. %
To show this, simply enumerate the ten first-order symmetries:
\begin{align}\label{eq:symmetry_operators}
  3 \text{\: translations \:} & \pop_j = \imagi \frac{\partial}{\partial x_j} \equiv \imagi \partial_j,\\ \nonumber
  3 \text{\: rotations \:} & \jop_j = \imagi \sum_{kl} \epsilon_{jkl} x_k \partial_l, \\ \nonumber
  3 \text{\: special conformal transformations \:} & \kop_j = \imagi \left[ \frac{5}{2}x_j + \sum_kx_jx_k\partial_k - \frac{1}{2} \sum_k x_k^2 \partial_j\right], \\ \nonumber
  \text{one~dilatation \:} & \dop = \imagi \left[ \frac{5}{2} + \sum_j x_j \partial_j \right].
\end{align}
We see that six of the operators are self-adjoint and commute with the
Laplacian
\begin{equation}
  [\pop_j, \Lap] = 0, \quad \pop_j^* = \pop_j, \qquad
  [\jop_j, \Lap] = 0, \quad \jop_j^* = \jop_j,
\end{equation}
and the remaining four are not self-adjoint but do quasi-commute,
\begin{align}
  \kop_j\Lap = \Lap\kop_j^*, & \qquad \kop_j^* = \kop_j - 2 \imagi x_j, \\ \nonumber
  \dop\Lap = \Lap\dop^*, & \qquad  \dop^* = \dop - 2 \imagi.
\end{align}
Finally, observe that quasi-commutation is preserved when taking
symmetric products of operators: given quasi-commuting $X$ and
$Y$, we have
\begin{equation}
\{X, Y\} \Lap - \Lap \{X, Y\}^* = XY\Lap + YX\Lap - \Lap Y^*X^* - \Lap X^*Y^* = 0,
\end{equation}
The remaining possibilities are $(Z\Lap)^* = \Lap Z^*$ for any
operator or function $Z$, and any constants. So the quasi-commuting
operators consist of \refedit{symmetric real-linear combinations} of
the algebra's first-order basis elements, plus any constants and terms
of the form $Z\Lap$.

\subsection{Disc operators}\label{sec:disc_operators}

On the disc $z=0$ \eqref{eq:operator_disc_mapping} the relevant new
operators are
\begin{align}
  \mathpzc{k}_1 = \imagi\left(\frac{3x}{2} + \frac{x^2-y^2}{2}\partial_x + xy\partial_y\right), \qquad \mathpzc{k}_1^* = \mathpzc{k}_1 - \imagi x, \\ \nonumber
  \mathpzc{k}_2 = \imagi\left(\frac{3y}{2} + \frac{y^2-x^2}{2}\partial_y + xy\partial_x\right), \qquad \mathpzc{k}_2^* = \mathpzc{k}_2 - \imagi y, \\ \nonumber
  \mathpzc{d} = \imagi \left( \frac{3}{2} + x\partial_x + y\partial_y\right), \qquad \mathpzc{d}^* = \mathpzc{d} - \imagi.
\end{align}

\section{Commutation of $T_j$}\label{sec:operator_deriv}

We must find a function $u$ such that the commutation relation
\begin{equation}\label{eq:t1_t2_commute}
  [T_1^*, T_2^*] = [g L_1, gL_2 ] + [g \Lap, u] = 0
\end{equation}
holds. In the simply separable case $u = 0$ suffices, but we need
$u \neq 0$ in $R$-separable coordinates.  Expanding out
\eqref{eq:t1_t2_commute} we find that in general we must have
\begin{equation}\label{eq:u_def_integral}
  u \propto \frac{1}{2} \int d q_2 \left( \partial_1 \log{\left(R^2 f_1\right)} \partial_1 \partial_2 \log{R^2} + \partial_1^2 \partial_2 \log{R^2} \right),
\end{equation}
or equivalently a similar expression with the two lower indices
interchanged. For the two $R$-separable coordinate systems considered
in this paper, this expression may be immediately integrated to yield
\eqref{eq:u_def}, and we can verify that this makes the commutator
\eqref{eq:t1_t2_commute} vanish in both cases. Additionally we find
that $\nabla^2 u = 0$ and $u \to 0$ at infinity.

However, for some more exotic coordinate systems we must use the
freedom to add an arbitrary function of $q_2$ to
\eqref{eq:u_def_integral} when integrating; for example, in cardioid
coordinates \citep[p.~107]{Moon1988} the correct definition is
\begin{equation}
  u \propto \frac{1}{2} \partial_1^2 \log{R^2} + \frac{1}{4} \left( \partial_1\log{\left(R^2 f_1\right)} \right)^2 - \frac{1}{2} q_2^{-2}.
\end{equation}
These definitions of $u$ are essentially ad hoc, and a sounder
theoretical basis would be desirable.

\section{Separable coordinates and conformal Killing tensors}\label{sec:Sj_Tj_relationship}

The classification of $R$-separable coordinate systems
\citep[Ch.~3]{Miller1984}\footnote{First published as
  \citet{Boyer1976}.} associates a pair of commuting quadratic
symmetries $(S_1,S_2)$ of the Laplacian with each of the $14$
canonical separable coordinate systems\footnote{Modulo an ambiguity
  discussed in \citet[\S 3.4.3]{Kalnins2018}.}. Given that these
operators $S_j$ are already known, it would be convenient if we could
construct our quasi-commuting operators $T_j$ out of them.

Quasi-commuting operators \eqref{eq:quasi} generically take the form
\begin{equation}\label{eq:generic_symmetry_operator}
  T^* = \mathcal{D}_K + A\Lap,
\end{equation}
where $K$ is a symmetric conformal Killing tensor (CKT), $A$ is a
self-adjoint operator, and $\mathcal{D}_K$ is a canonical map between
CKTs and differential operators defined in
\citet[Eq.~(5)]{Eastwood2002}\footnote{Eastwood's maps
  $(\mathcal{D}_K,\delta_K)$ satisfy
  $\Lap \mathcal{D}_K = \delta_K \Lap$ for any CKT $K$; the algebraic
  structure is obtained by considering two generalised symmetries to
  be equivalent if they differ by a term of the form $A\Lap$, and
  hence each symmetry operator is canonically associated with a single
  CKT.}. For $T$ a second-order operator we have $K$ a rank-2 CKT and
$A$ a zeroth-degree operator (a function). \refedit{If we write
  $S_j = \mathcal{D}_{Q_j}$ ($j = 1,2$) for the pair of operators
  associated with some $R$-separable coordinate system, and
  $T_j^* = \mathcal{D}_{K_j} + f_j\Lap$ ($j = 1,2,3$) for our desired
  triplet of commuting operators, then we could write a `trivial'
  ansatz for the $T_j$ by making each $K_j$ a linear combination of
  the \emph{same} pair of $Q_j$, and afterwards compute such $f_j$
  that account for the remaining terms in the commutation relation
  (proportional to $\Lap$)} \footnote{\refedit{Commuting two generic
    second-order symmetry operators of the form
    \eqref{eq:generic_symmetry_operator} using the composition rule
    given in \citet[Eq.~(30)]{Eastwood2002}, we find
    \[ [\mathcal{D}_{K_1} + f_1\Lap, \mathcal{D}_{K_2} + f_2\Lap] =
      \mathcal{D}_{[K_1,K_2]} + \mathcal{D}_{X(K_1,K_2)} +
      Y(K_1,K_2,f_1,f_2)\Lap = 0, \] where $[K_1, K_2]$ is the
    (rank-3) Schouten-Nijenhuis bracket, $X(K_1,K_2)$ is a rank-1 CKT
    that depends linearly on each component of the $K_j$, and
    $Y(K_1,K_2,f_1,f_2)$ is a first-order differential operator. The
    spaces of rank-$(1,2,3)$ CKTs have respectively $(10,35,84)$
    dimensions, so solving the CKT part of the commutation relation to
    find the $K_j$ involves an overdetermined system of $84+10=94$
    bilinear equations in $2 \cdot 35 = 70$ variables (this
    computation is avoided if the simplifying ansatz described above
    is used). A system of two second-order PDEs must then be solved to
    find the $f_j$.}}. \refedit{In fact this `trivial' approach
  accounts for all the $T_j$ considered in the present work, and on
  heuristic grounds it seems likely that this approach will also work
  for the remaining separable coordinate systems.}

Choosing the CKT part of each $T_j$ according to the ansatz above, we
have\footnote{\refedit{The optional constants $c_j$ may be introduced
    to simplify certain expressions involving eigenvalues.}}
\begin{equation}\label{eq:T1T2S1S2_relation}
  T_j^* = a_j S_{1} + b_j S_{2} + c_j + h_j\Lap, \quad (j = 1,2,3)
\end{equation}
for some constants $(a_j,b_j,c_j)$ and functions $h_j$.
Then by Gaussian elimination we can find constants
$d_j = (\vec{a} \times \vec{b})_j$ and a function
$g = -\vec{d} \cdot (\vec{h} + \vec{c})$ such that
\begin{equation}\label{eq:T1T2T3_laplacian_relation}
  d_1 T_1^* + d_2 T_2^* + d_3 T_3^* = -g \nabla^2.
\end{equation}
This is just the generalisation of \eqref{eq:t1_t2_lap} to any
$R$-separable coordinate system.
This implies the simple relationship
\eqref{eq:eigen_potential_density_relation} between eigen-potentials
and eigen-densities, but does not uniquely specify $T_j$ or $g$. In
fact the original ansatz \eqref{eq:Tj_def} is just one possible way of
defining a set of commuting $T_j$ (for example,
\eqref{eq:spherical_alternate} is an alternative solution in spherical
polar coordinates). \refedit{Note that for rotational
  (resp.~cylindrical) coordinates the simplification is due to having
  $T_3 = S_2 = \mathcal{J}_3^2$ (resp.~$\mathcal{P}_3^2$) in
  \eqref{eq:T1T2S1S2_relation} and hence $d_3 = 0$ in
  \eqref{eq:T1T2T3_laplacian_relation}.}

As an aside we observe that from \eqref{eq:T1T2T3_laplacian_relation}
we can find the familiar separable solutions to Laplace's equation
from the eigen-potentials $\phi_{\lambda}$ by choosing a vector of
eigenvalues $\lambda$ that lies in the plane
$\lambda \cdot \vec{A} = 0$ (in the rotational or cylindrical cases
simply $\lambda_1 = -d_2 \lambda_2 / d_1$). In this way we can view our
results as a kind of generalisation of the classical theory of
separation of variables for Laplace's equation to cover the
inhomogeneous case, i.e.~Poisson's equation.

Finally we note that the method of Helmholtz wavefunctions discussed
in Sec.~\ref{sec:intro} is equivalent to making the choices
$T_1 = \Lap$, $T_2 = S_1$, $T_3 = S_2$ in
\eqref{eq:T1T2S1S2_relation}; \refedit{in this case we always have
  $T_j^* = T_j$, as the Helmholtz equation only separates in the
  simply separable coordinates, so $\dop$ and $\kop_j$ never appear in
  any of the $S_j$}.

\section{Further prolate spheroidal basis sets}\label{sec:more_prolate_spheroidal}

Following Sec.~\ref{sec:spheroidal_plummer}, we can try a more general
functional form for the zeroth-order potential by introducing a
dimensionless `scale-length' $a$,
\begin{equation}\label{eq:spheroidal_generalised_potential}
  \Phi_{0lm}^{(a)}(\eta,\vartheta,\varphi) = \frac{-\sinh^{|m|}{\!\eta}}{(a + \cosh\eta)^{l+|m|+1}} \: P_l^{(m)}(\cos\vartheta) \: \expe^{\imagi m \varphi}.
\end{equation}
We can perform the Mehler-Fock transform \prud{Vol.~3, 2.17.4(9)} and
find the weight function,
\begin{equation}\label{eq:spheroidal_generalised_weight}
  \omega_{lm}^{(a)}(\alpha^2) = \frac{(a^2-1)^{-l}}{4 (l+|m|)!^2}
  \frac{\alpha \sinh{\!(\pi\alpha)}}{\cosh{\!(\pi\alpha)}^2} \left| (l + 1/2 + \imagi \alpha) \: \Gamma(|m|+1/2+\imagi\alpha) \: P_{\imagi \alpha - 1/2}^{(l)}\!\left(a\right) \right|^2.
\end{equation}
This has a non-trivial dependence on $a$, so we would not expect it to
simplify in general, but due to the limiting behaviour of the Legendre
function \dlmf{14.8} there are two special cases: at $a = 0$ we
recover the `Plummer/scale-free' case of
Sec.~\ref{sec:spheroidal_plummer}; and at $a=1$ we find the density
simplifies to
\begin{equation}
  \varrho_{0lm}^{(1)}(\eta,\vartheta,\varphi) = \frac{(1+l)(1+l+|m|)}{2\pi b^2} \: \frac{\sinh^{|m|}{\!\eta}}{(\sinh^{2}\!\eta + \sin^{2}\!\vartheta) \: (1 + \cosh\eta)^{l+|m|+2}}
  \: P_l^{(m)}(\cos\vartheta) \: \expe^{\imagi m \varphi},
\end{equation}
and the weight function to
\begin{equation}\label{eq:pro_sph_iso_weight}
  \omega_{lm}^{(1)}(\alpha^2) = \frac{1}{\pi^2 2^{2l+4} \left(l! (l+|m|)!\right)^2}
  \left| \frac{\Gamma(|m|+1/2+\imagi\alpha) \: \Gamma(l+1/2+\imagi\alpha) \: \Gamma(l+3/2+\imagi\alpha) \: \Gamma(1/2+\imagi\alpha)}{\Gamma(2 \imagi \alpha)} \right|^2,
\end{equation}
which is proportional to the weight for the Wilson polynomials
$W_n(\alpha^2;a_m,b_l,c_l,d)$ with parameters
$(a_m,b_l,c_l,d) = (|m|+1/2, l+1/2, l+3/2, 1/2)$. So we take this as our
index-raising polynomial, and find that the corresponding potential
and density functions are again proportional to Jacobi polynomials,
\begin{align}\label{eq:spheroidal_isochrone_basis}
  \Phi_{nlm}^{(1)}(\eta,\vartheta,\varphi) &= A_{nlm} \: \Phi_{0lm}^{(1)}(\eta,\vartheta,\varphi) \: P_n^{(2l+1,|m|)}\!\!\left(\zeta\right), \\ \nonumber
  \varrho_{nlm}^{(1)}(\eta,\vartheta,\varphi) &= A_{n,l+1,m} \:  \varrho_{0lm}^{(1)}(\eta,\vartheta,\varphi) \: P_n^{(2l+1,|m|)}\!\!\left(\zeta\right), \\ \nonumber
  \zeta = \frac{\sinh^{2}\!\left(\eta/2\right) - 1}{\sinh^{2}\!\left(\eta/2\right) + 1}, \quad&\quad A_{nlm} =  n! (l+1)_n (l+|m|+1)_n.
\end{align}
On the plane $z=0$ at zeroth order this reproduces the isochrone
model, but away from the plane it is not in Stäckel form. Comparison
of the $a=0$ \eqref{eq:spheroidal_plummer_basis} and $a=1$
\eqref{eq:spheroidal_isochrone_basis} cases would seem to suggest a
simple formula for the intermediate values of $a$, but this is stymied
by the complicated form of \eqref{eq:spheroidal_generalised_weight}
(of course, the intermediate cases can be produced numerically).

\begin{figure}
  \centering
  \includegraphics[width=\textwidth]{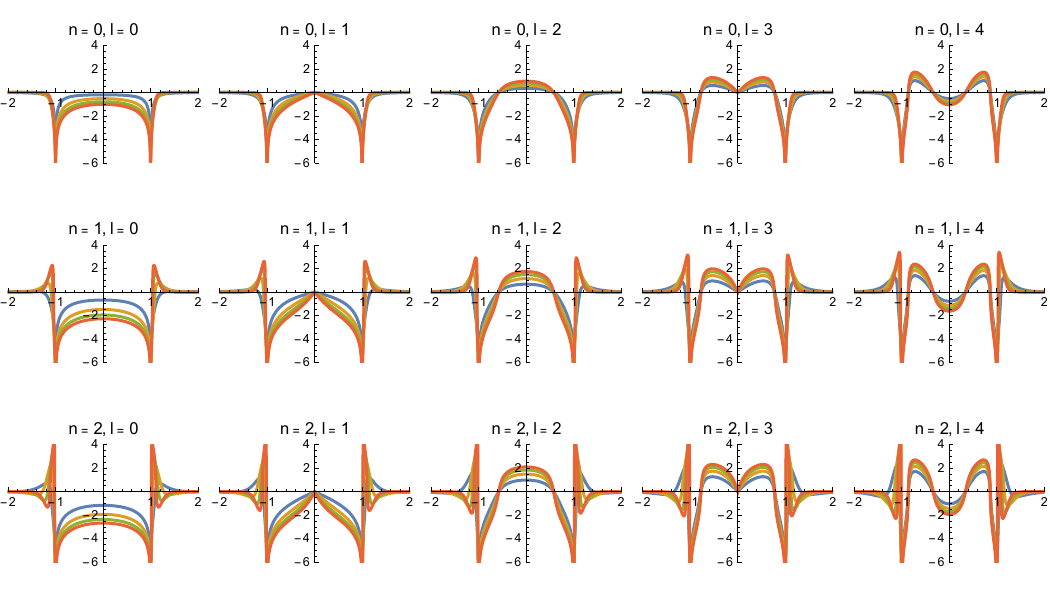}
  \caption{\refedit{The result \eqref{eq:prolate_operator_identity_0}
      allows us to semi-numerically produce perturbations to the
      `Plummer/scale-free' basis set of
      Sec.~\ref{sec:spheroidal_plummer} such that each density basis
      function contains an additional factor of
      $\cosh^{-2q}\eta$. Here we show density functions for
      $q \in (0,3,6,9)$, $n = 0\ldots 2$, $l = 0\ldots 4$, scaled and
      normalised by
      $\arsinh{\!\left(\varrho_{nl0}(0,0,z)/\sqrt{N_{nl0}}\right)}$,
      with $z$-axis horizontal and
      $x=y=0$.}}\label{fig:spheroidal_plummer_q}
\end{figure}

In fact, both the $a=0$ and $a=1$ cases could have been derived by
making a suitable variable substitution ($\xi$ or $\zeta$) in
Poisson's equation, similar to the derivations of \citet{CB73,Zh96},
but this is not possible when taking slightly more general forms for
the zeroth-order density. For example, operating on
$\varrho_{0lm}^{(1)}$ with a certain polynomial in $T_1$ leads to a
family of generalised power-law densities (with parameter $q$)
\begin{equation}\label{eq:prolate_operator_identity_1}
  \left| \left(l + 3/2 + \imagi \sqrt{T_1}\right)_q \right|^2 \varrho_{0lm}^{(1)} =
  \frac{2^q \: (l+1)_{q+1} \: (l+|m|+1)_{q+1} \: \sinh^{|m|}{\!\eta} \: Y_{lm}(\vartheta,\varphi)}{2 \pi b^2 \: (1 + \cosh\eta)^{l+|m|+2+q} \: (\sinh^{2}\!\eta + \sin^{2}\!\vartheta)},
\end{equation}
which gives a weight function similar to \eqref{eq:pro_sph_iso_weight}
but containing an additional factor of
$| (l + 3/2 + \imagi \alpha)_q |^4$; applying a similar operator to
$\varrho_{0lm}^{(0)}$
leads to a family of `scale-free' power-law densities
\begin{equation}\label{eq:prolate_operator_identity_0}
  \left| \left(l/2 + 5/4 + \imagi \sqrt{T_1}/2\right)_q \right|^2 \varrho_{0lm}^{(0)} =
  \frac{2 \left(\left(l+|m|+1\right)/2\right)_{q+1} \left(\left(l+|m|+2\right)/2\right)_{q+1} \sinh^{|m|}{\!\eta} \: Y_{lm}(\vartheta,\varphi)}{\pi b^2 \: \cosh^{l+|m|+3+2q}{\!\eta} \left(\sinh^2\!\eta + \sin^2\!\vartheta\right)},
\end{equation}
whose weight function is similar to
\eqref{eq:spheroidal_plummer_weight} but contains an additional factor
of~\mbox{$| (l/2 + 5/4 + \imagi \alpha/2)_q |^4$}. Neither weight
function appears to give classical polynomials, but the simple form
ameliorates their numerical construction. In
Fig.~\ref{fig:spheroidal_plummer_q} we plot some examples of density
basis functions with varying $q$ constructed using the modified
Chebyshev algorithm \citep{Gautschi1990}.

Another useful result arises from using
\eqref{eq:spheroidal_generalised_potential} as the zeroth-order
\emph{density}, and then taking the limits $a,l \to \infty$; we obtain
the exponentially-declining form
\begin{equation}
  \varrho_{0lm}^{(p)}(\eta,\vartheta,\varphi) \propto \frac{\sinh^{|m|}{\!\eta} \: \expe^{-p\cosh\eta }}{\sinh^{2}\!\eta + \sin^{2}\!\vartheta} \: P_l^{(m)}(\cos\vartheta) \: \expe^{\imagi m \varphi},
\end{equation}
whose weight function is \prud{Vol.~3, 2.17.7(5)}
\begin{equation}
  \omega_{m}^{(p)}(\alpha^2) \propto \alpha \sinh{\!(\pi\alpha)} \:  \left| \Gamma(|m|+1/2+\imagi\alpha) \: K_{\imagi\alpha}(p)\right|^2.
\end{equation}
This basis set then has roughly the same asymptotic behaviour as
prolate spheroidal \refedit{diatomic Hylleraas} wavefunctions \citep{Mendl2012},
although again the polynomials would have to be determined
numerically.

\section{Other cylindrical coordinate systems}\label{sec:other_cylindrical}

\subsection[Elliptic cylinder]{Elliptic cylindrical \coordref{III}{4}}
\begin{flalign*}
\text{\textit{Coords:}} \quad & q_1 = \eta, \quad q_2 = \varphi, \quad q_3 = z, \quad x = b\cosh\eta \cos\varphi, \quad y = b\sinh\eta\sin\varphi && \\
                              & h_1 = -h_2 = b\sqrt{\cosh^{2}\!\eta - \cos^{2}\varphi}, \quad f_1 = -f_2 = \refedit{-h_3} = R = 1, \quad g = b^2\left(\cos^{2}\!\varphi - \cosh^{2}\!\eta\right) && \\
\text{\textit{Operators:}} \quad & L_1 = -g^{-1}\partial_\eta^2, \quad L_2 = -g^{-1}\partial_\varphi^2, \\
                              & \refedit{T_1^* = -\jop_3^2 - b^2\pop_1^2 - b^2\cosh^{2}\!\eta \Lap, \quad T_2^* = \jop_3^2 + b^2\pop_1^2 + b^2\cos^{2}\!\varphi \Lap, \quad T_3 = \pop_3}
\end{flalign*}
As in the cylindrical polar case we exchanged the roles of the
$\varphi$ and $z$ coordinates. The eigen-potentials and densities are
\begin{align}
  \phi_{\alpha m k}(\eta,\varphi,z) &= \frac{-4 \pi}{\alpha^2 + m^2} \: \expe^{\imagi k z} \: \me_{2q + \alpha^2}(\imagi \eta - \pi/2, q) \: \me_{2q - m^2}(\varphi - \pi/2, q) , \\ \nonumber
  \Psi_{\alpha m k}(\eta,\varphi,z) &= \frac{b^{-2}}{\cos^{2}\!\varphi - \cosh^{2}\!\eta} \: \expe^{\imagi k z} \: \me_{2q + \alpha^2}(\imagi \eta - \pi/2, q) \: \me_{2q - m^2}(\varphi - \pi/2 , q),
\end{align}
where $q = \left(bk/2\right)^2$ and
$\me_\nu(z,q) = \ce_\nu(z,q) + \imagi \se_\nu(z,q)$ is (generically) a
complex Mathieu function. The corresponding integral transform has
been studied in \citet{Habashy1986,Naylor1989,InayatHussain1991}.

\subsection[Parabolic cylindrical]{Parabolic cylindrical \coordref{IV}{3}}
\begin{flalign*}
\text{\textit{Coords:}} \quad & q_1 = \lambda, \quad q_2 = \mu, \quad q_3 = z, \quad x = (\lambda^2-\mu^2)/2, \quad y = \lambda\mu && \\
                              & h_1 = h_2 = \sqrt{\lambda^2+\mu^2}, \quad f_1 = f_2 = \refedit{-h_3} = R = 1, \quad g = -(\lambda^2+\mu^2) && \\
\text{\textit{Operators:}} \quad & L_1 = -g^{-1}\partial_\lambda^2, \quad L_2 = -g^{-1}\partial_\mu^2, \\
                              & \refedit{T_1^* = \{\jop_3, \pop_2\} - \lambda^2\Lap, \quad T_2^* = -\{\jop_3, \pop_2\} - \mu^2\Lap, \quad T_3 = \pop_3}
\end{flalign*}
Another cylindrical system. The eigenfunctions are
\begin{align}
  \phi_{\alpha\beta k}(\lambda,\mu,z) &= \frac{-4 \pi}{\alpha^2 + \beta^2} \: \expe^{\imagi k z} \: U_{\alpha^2/(2k)}\!\left(\sqrt{2k}\lambda\right) \: U_{\beta^2/(2k)}\!\left(\sqrt{2k}\mu\right), \\ \nonumber
  \Psi_{\alpha\beta k}(\lambda,\mu,z) &= \frac{-1}{\lambda^2+\mu^2} \: \expe^{\imagi k z} \: U_{\alpha^2/(2k)}\!\left(\sqrt{2k}\lambda\right) \: U_{\beta^2/(2k)}\!\left(\sqrt{2k}\mu\right),
\end{align}
where $U_a(z)$ is a parabolic cylinder function \dlmf{12}; however the
integral transform apparently does not exist for unrestricted
$\lambda$ and $\mu$ \citep{Jerri1982}.

\end{document}